\documentstyle[12pt]{article}
\def\be{\begin{equation}}
\def\ee{\end{equation}}
\def\bea{\begin{eqnarray}}
\def\eea{\end{eqnarray}}
\def\beas{\begin{eqnarray*}}
\def\eeas{\end{eqnarray*}}
\def\no{\nonumber}
\def\a{\alpha}

\def\d{\delta}

\def\m{\mu}
\def\n{\nu}
\def\LLL{\Lambda}
\def\lll{\lambda}

\def\O{\Omega}
\def\p{\phi}

\def\ps{\psi}
\def\r{\rho}

\def\bra{\langle}
\def\ket{\rangle}
\newcommand{\lbra}{\left\langle}
\newcommand{\rket}{\right\rangle}

\def\eq{\equiv}

\def\pa{\partial}
\def\ra{\rightarrow}

\def\intdx{\int\! d^4\!x\,}
\newcommand{\bbR}{{\sf R\hspace*{-0.9ex}\rule{0.15ex}{1.5ex}\hspace*{0.9ex}}}

\def\lag{{\cal L}}
\newcommand{\np}[1]{Nucl.\ Phys.\ {\bf {#1}}}
\newcommand{\plt}[1]{Phys.\ Lett.\ {\bf {#1}}}

\newcommand{\prlt}[1]{Phys.\ Rev.\ Lett.\ {\bf {#1}}}
\newcommand{\ijmp}[1]{Int.\ J.\ Mod.\ Phys.\ {\bf {#1}}}
\newcommand{\rmps}[1]{Rev.\ Mod.\ Phys.\ {\bf {#1}}}
\newcommand{\prp}[1]{Phys.\ Rep.\ {\bf {#1}}}
\newcommand{\anp}[1]{Ann.\ Phys. (N. Y.)\ {\bf {#1}}}
\newcommand{\cmp}[1]{Comm.\ Math.\ Phys.\ {\bf {#1}}}

\renewcommand{\theequation}{\thesection.\arabic{equation}}
\newcommand{\lo}{{\Lambda_0}}
\newcommand{\dph}{{\cal D}\phi}
\newcommand{\dvl}[2]{{\cal D}_{#1,#2}^{\LLL}}
\newcommand{\dvli}[2]{{\cal D}_{#1,#2}^{\LLL;I}}
\newcommand{\dvln}[2]{{\cal D}_{#1,#2}^{\LLL;N}}
\newcommand{\dvlg}[3]{{\cal D}_{#2,#3}^{\LLL;#1}}
\newcommand{\dvlo}[2]{{\cal D}_{#1,#2}^{\lo}}
\newcommand{\dvlon}[2]{{\cal D}_{#1,#2}^{\lo;N}}
\newcommand{\dvo}[2]{{\cal D}_{#1,#2}^{\LLL=0}}
\newcommand{\dvon}[2]{{\cal D}_{#1,#2}^{\LLL=0;N}}
\newcommand{\dvll}[2]{{\cal D}_{#1,#2}^{\lambda}}
\newcommand{\dvllg}[3]{{\cal D}_{#2,#3}^{\lambda;#1}}
\newcommand{\lfilo}{L^{(f)\lo}}
\newcommand{\lfil}{L^{(f)\Lambda}}
\newcommand{\litlo}{L_{\scriptstyle{\rm tot}}^{\lo}}
\newcommand{\litl}{L_{\scriptstyle{\rm tot}}^{\Lambda}}
\newcommand{\lilo}{L^{\lo}}
\newcommand{\lil}{L^{\Lambda}}

\newcommand{\lilon}{L^{\lo;N}}
\newcommand{\lilog}[1]{L^{\lo;#1}}

\newcommand{\liln}{L^{\Lambda;N}}
\newcommand{\lilg}[1]{L^{\Lambda;#1}}

\newcommand{\ltilog}[1]{\widetilde L^{\lo;#1}}

\newcommand{\lfl}[2]{{\cal L}_{#1,#2}^{(f)\LLL}}
\newcommand{\lflo}[2]{{\cal L}_{#1,#2}^{(f)\lo}}
\newcommand{\lfo}[2]{{\cal L}_{#1,#2}^{(f)\LLL=0}}

\newcommand{\lfll}[2]{{\cal L}_{#1,#2}^{(f)\lambda}}
\newcommand{\lvl}[2]{{\cal L}_{#1,#2}^{\LLL}}
\newcommand{\lvlo}[2]{{\cal L}_{#1,#2}^{\lo}}
\newcommand{\lvlon}[2]{{\cal L}_{#1,#2}^{\lo;N}}

\newcommand{\lvo}[2]{{\cal L}_{#1,#2}^{\LLL=0}}

\newcommand{\lvon}[2]{{\cal L}_{#1,#2}^{\LLL=0;N}}

\newcommand{\lvog}[3]{{\cal L}_{#2,#3}^{\LLL=0;#1}}

\newcommand{\lvli}[2]{{\cal L}_{#1,#2}^{\LLL;I}}
\newcommand{\lvln}[2]{{\cal L}_{#1,#2}^{\LLL;N}}
\newcommand{\lvlg}[3]{{\cal L}_{#2,#3}^{\LLL;#1}}

\newcommand{\lvpl}[2]{{\cal L'}_{#1,#2}^{\LLL}}

\newcommand{\lvpli}[2]{{\cal L'}_{#1,#2}^{\LLL;I}}
\newcommand{\lvpln}[2]{{\cal L'}_{#1,#2}^{\LLL;N}}
\newcommand{\lvplg}[3]{{\cal L'}_{#2,#3}^{\LLL;#1}}

\newcommand{\ltog}[3]{\widetilde{\cal L}_{#2,#3}^{\LLL=0;#1}}
\newcommand{\lttog}[2]{\widetilde{\cal L}_{#2}^{\LLL=0;#1}}

\newcommand{\ltln}[2]{\widetilde{\cal L}_{#1,#2}^{\LLL;N}}
\newcommand{\ltlg}[3]{\widetilde{\cal L}_{#2,#3}^{\LLL;#1}}

\newcommand{\ltpln}[2]{\widetilde{\cal L'}_{#1,#2}^{\LLL;N}}
\newcommand{\ltplg}[3]{\widetilde{\cal L'}_{#2,#3}^{\LLL;#1}}
\newcommand{\lbl}[2]{\bar{\cal L}_{#1,#2}^{(f)\LLL}}
\newcommand{\lbo}[2]{\bar{\cal L}_{#1,#2}^{(f)\LLL=0}}

\newcommand{\fvln}[2]{{\cal F}_{#1,#2}^{\LLL;N}}

\newcommand{\fvlln}[2]{{\cal F}_{#1,#2}^{\lll;N}}

\newcommand{\dlo}{{P^{\lo}}}
\newcommand{\dl}{{P^{\Lambda}}}
\newcommand{\dllo}{{P_{\Lambda}^{\lo}}}
\newcommand{\dmlo}{{P_m^{\lo}}}
\newcommand{\dml}{{P_m^{\Lambda}}}

\newcommand{\dMlo}{{P_M^{\lo}}}
\newcommand{\dMl}{{P_M^{\Lambda}}}

\newcommand{\taylor}{\tau^{4+2N-2n}}
\newcommand{\pal}{\pa_\Lambda}
\newcommand{\palo}{\pa_\lo}
\newcommand{\plog}[2]{{\rm Plog}\left(\frac{#1}{#2}\right)}
\newcommand{\norm}[2]{\left\|{#1}\right\|_{#2}}
\newcommand{\norml}[1]{\left\|{#1}\right\|_{(2\LLL,\m)}}
\newcommand{\normn}[1]{\left\|{#1}\right\|_{(2\LLL,\m,M)}}
\newcommand{\intdp}{\int\frac{d^4p}{(2\pi)^4}}
\newcommand{\intdq}{\int\frac{d^4q}{(2\pi)^4}}
\newcommand{\mono}[2]{M_{#1,\{#2\}}}
\newcommand{\caln}{{\cal N}_{\{w\}}}
\newcommand{\qedbox}{\vrule height 8pt width 5pt depth 1pt}
\newtheorem{theorem}{Theorem}
\begin{document}
\begin{flushright}SNUTP 94-20\end{flushright}
\begin{flushright}hep-th/9402117\end{flushright}

\vskip 1cm

\begin{center}
{\Large\bf A Renormalization Group Flow Approach to\\[5mm]
\mbox{Decoupling and Irrelevant Operators}}

\vskip 1.5cm
Chanju Kim$^*$\\[2mm]

{\it Department of Physics and Center for Theoretical Physics,\\
Seoul National University, Seoul, 151-742, Korea}
\end{center}
\vspace{12mm}
\begin{center}
{\large\bf Abstract}\\[5mm]
\end{center}
\ \indent
Using Wilson-Polchinski renormalization group equations, we give a
simple new proof of decoupling in a $\p^4$-type scalar field theory involving
two real scalar fields (one is heavy with mass $M$ and the other light).
Then, to all orders in perturbation theory, it is shown that effects
of virtual heavy particles up to the order $1/M^{2N_0}$ can be systematically
incorporated into light-particle theory via
effective local vertices of canonical dimension at most $4+2N_0$.
The couplings for
vertices of dimension $4+2N$ are of order $1/M^{2N}$ and are systematically
calculable. All this is achieved through intuitive dimensional arguments
without resorting to complicated graphical arguments or convergence theorems.

\vspace{40mm}
\hrule width 50mm
\vspace{5mm}
{\footnotesize * e-mail: cjkim@phyb.snu.ac.kr}

\newpage
\section{INTRODUCTION}
\ \indent
The concept of effective field theories has played an important role in modern
theoretical physics
and it acquires its natural physical
interpretation in the Wilson renormalization group formalism \cite{wilson}.
In the latter, one integrates out the high
frequency modes scanned by a
cutoff
and then considers lowering the
cutoff.
This generates the RG flow, and
the differential equation governing this flow is called
the exact renormalization group equation or the flow equation. In general, as
we scale down to a lower cutoff, the (Wilsonian) effective Lagrangian
converges toward a finite-dimensional manifold parametrized by the
``relevant'' couplings which correspond to the renormalizable couplings in the
conventional perturbation theory.
This explains the phenomenological success of renormalizable
quantum field theories in particle physics. Polchinski was able to
give a simple proof of the
perturbative renormalizability
within this framework \cite{polchinski}, taking the $\p^4$-theory as his model.
Since the idea is very general, it
has been extended to many different field-theory models, to study
renormalization of composite
operators and short-distance expansions, and also toward more practical
applications. See for example \cite{warr}--\cite{ball}.

If we take seriously the above point of view on effective field theory and
renormalization, it should be natural also
to find, using this flow equation approach, an intuitive, yet rigorous,
proof of the decoupling theorem \cite{applequist} which states that,
in a generic renormalizable quantum field theory with heavy particles
of mass $M$, heavy-particle effects
decouple from low-energy light-particle physics except for renormalization
of couplings involving light fields and corrections of order $1/M^2$.
In fact, as far as low-energy physics is concerned, heavy fields are not very
different from massive regulator fields and hence those fields may be
integrated out \cite{weinberg} (together with high-frequency modes for light
fields) as one lowers the cutoff below the heavy mass scale $M$. One then
expects that
the nonlocal irrelevant terms,
generated in the effective Lagrangian by this integrating-out procedure, be
suppressed as the cutoff is scaled further down, assuming that
they are well-behaved. In this light,
decoupling looks quite natural. Indeed, the decoupling theorem can be proved
precisely along this line, and in this approach complicated graphical
arguments or convergence theorems are entirely dispensed with.

Since the resulting
effective theory is renormalizable
to the zeroth order in $1/M^2$,
if
one wishes to understand the low-energy
manifestations of heavy particles,
irrelevant
(nonrenormalizable) terms must be considered.
This issue was investigated in Refs.\ \cite{clee,kazama} and
the result is that
virtual heavy particle effects can be isolated via a set
of effective {\it local} vertices with calculable couplings
when combined with appropriate calculation rules to
deal with irrelevant vertices.
This {\it factorization}
theorem was proved to order $1/M^2$ by adapting Zimmermann's algebraic
identities \cite{zimmermann} in the BPHZ formulation. The proof involves
elaborate mathematical rearrangement and convergence theorems, and thus
looks a bit artificial. [Also, as for the subgraphs involving heavy and light
particle legs simultaneously, the precise nature of factorization is not very
clear in this approach.]
In this paper, we present a rather elementary proof of this heavy-mass
factorization
to all orders
in perturbation theory
and
to any given order in
$1/M^2$, using
flow equations.
Actually
the whole scheme allows a natural physical interpretation from the viewpoint
of the RG
flow.
The calculation rules for irrelevant vertices are also systematically given.

There is, however, a subtle point.
Our hope is
to find a local
effective theory with
the UV cutoff $\lo$ ($\gg M$)
which is accurate
up to the order $1/M^{2N_0}$ ($N_0$ is any fixed integer).
In the original Wilson's view, the cutoff scale of the effective
theory may well be identified with the heavy mass scale $M$, above
which it is no more effective,
and there is no need to worry about the
presence
of irrelevant terms in the Lagrangian
in particular,
since
``natural scale'' of those terms will be around $M$.
Among them we may choose to keep explicitly some
minimal number of irrelevant terms
in our effective Lagrangian\footnote{For example,
the continuum version of Symanzik's improved action
\cite{symanzik} in lattice theory have been discussed in \cite{keller4}
by adding suitable irrelevant terms in Lagrangian.}
for the accuracy of order $1/M^{2N_0}$.
But, in the conventional discussion of quantum field
theory, the UV cutoff $\lo$ is supposed to go to infinity eventually.
So, to connect
it with
Wilson's view,
we may
suppose scaling up the cutoff of the above Wilsonian effective Lagrangian
from $M$ to
$\lo$. It will then generate infinitely many irrelevant terms which are
unnaturally large. Also, during the scaling, all of them
get mixed together and
so we are forced to work with the Lagrangian consisting of infinitely many
terms all the time. Any kind of truncation for the bare Lagrangian to some
finite number of terms would yield divergences in physical quantities as
$\lo\ra\infty$, because the unnaturally large coefficients would be amplified
by some positive power of $\lo/M$ as the cutoff is scaled down.
This is nothing but the statement of nonrenormalizability in the
language of RG
flow. To avoid this problem we need to deal with irrelevant terms
carefully, i.e., give suitable rules to obtain unambiguous finite results with
only a finite number of terms included in the bare Lagrangian. It is
achieved through the modification of the flow equation when irrelevant
vertices are inserted, in the more-or-less same way as one treats
composite operators
and their normal products \cite{zimmermann} in the
flow equation approach \cite{keller}.
Based on this procedure,
we can connect the
BPHZ approach
with the present one.

The plan of our paper is as follows. In Sec.\ 2, we present some basics
on the flow equation, which are used throughout this paper.
In Sec.\ 3, we define our model theory (the full theory)
and choose a suitable form
of cutoff functions which implement naturally the idea of effective field
theory discussed above. We then obtain certain bounds on Green functions which
are useful in the next section
and prove the perturbative
renormalizability. In Sec.\ 4, we prove the decoupling theorem and then
proceed to show
the factorization of virtual heavy-particle effects by deriving
the local low-energy effective field theory which describes
low-energy light-particle physics to any desired order in powers of $1/M^2$.
We conclude in Sec.\ 5. In the Appendix, we briefly describe
the renormalization of Green functions with single or multiple insertion of
local vertices and Zimmermann's normal product which are needed in Sec.\ 4.

\setcounter{equation}{0}
\section{THE FLOW EQUATION}
\ \indent
In this section we present some features of the flow equation in
general setting to fix notations and to facilitate our later analysis.
For simplicity we consider a theory of a single scalar
field in four Euclidean dimensions with a
cutoff $\lo$.
The bare action is written as
\be
S^{\lo}[\phi]=\frac12\lbra\p,\dlo^{-1}\p\rket+L^{\lo}[\p]\,,
\ee
where $\dlo$ is the free-particle cutoff propagator,
$\bra f,g\ket$ is
defined by the momentum-space integral of $f$ and $g$,
$ \bra f,g\ket\equiv\intdp f(p)g(-p)\,, $
and $\lilo[\p]$ represents the interaction part of the bare action.
We will also allow the insertions of
some additional local vertices or certain composite operators (of unspecified
physical origin). To account for this, we define
$\litlo[\p]$ as a formal power series in $\a\eq(\a_1,\ldots,\a_k)$,
which has $\lilo[\p]$ as its zeroth
order term, viz.,
\be \label{litlo}
\litlo[\p]
  =\sum_{|N|\ge0}\frac{\a^N}{N!}\lilon[\p]\,,\quad (\lilo[\p]\eq\lilog{0}[\p])
\ee
where $N$ is a multiindex $N\eq(N_1,\ldots,N_k)$, $|N|\eq\sum_{i=1}^kN_i$,
$N!\eq N_1!\cdots N_k!$, and $\a^N\eq\a_1^{N_1}\cdots\a_k^{N_k}$.
$\lilon$'s may be regarded as additional local vertices appended to the
original Lagrangian $\lilo$ or as composite operators in which one is
interested.
Also,
we will denote
\be
L_C^\lo[\p]=\sum_{|N|\ge1}\frac{\a^N}{N!}\lilon[\p]
        \qquad\left(=\litlo[\p]-\lilo[\p]\right)\,.
\ee

The generating functional, with the insertion of the operator
$e^{L_C^\lo[\p]}$, is
\be\label{zj}
Z[J]=\int\dph e^{-\frac12\bra\p,\dlo^{-1}\p\ket-\litlo[\p]+\bra J,\p\ket}.
\ee
Following Wilson and Polchinski \cite{wilson,polchinski},
one may
integrate out the high-momentum components of $\p$
and reduce the cutoff $\lo$ to a lower scale $\LLL$.
The result is \cite{keller,morris}
\begin{equation} \label{lowz}
Z[J]=\int\dph e^{-\frac12\bra\p,\dl^{-1}\p\ket-\litl[\dllo J+\p]
                 +\bra J,\p\ket+\frac12\bra J,\dllo J\ket}\,,
\end{equation}
where $\dl$ ($\dllo$) is the low-(high-)frequency part of the
propagator\footnote{There
will be infinite ways in doing the separation; choosing one way of
separation may be considered as choosing a ``renormalization
scheme'' in the Wilson renormalization group approach,
and physical quantities are independent of such choices. While
it is not necessary to specify a particular scheme at this stage, we will
choose a specific scheme later to facilitate our discussions.}
with $\dlo=\dl+\dllo$, and
$\litl$ is
the generating functional of amputated connected Green
functions (with
$e^{L_C^\lo[\p]}$ inserted) with
{\it both\/} UV cutoff $\lo$ and IR cutoff $\LLL$.
Physical Green functions are obtained from $\litl$ in the limit $\LLL\ra0$
\cite{keller,bonini,morris}.
On the other hand, if $J(p)=0$ for $p>\LLL$ so that $J$ couples to the
low-frequency modes only, all $J$'s drop out from Eq.\ (\ref{lowz})
except for $\bra J,\p\ket$ since $\dllo$ has only high-frequency modes,
and $\litl$ coincides with the Wilsonian effective action \cite{wilson}.
$\litl$ obeys the exact renormalization group equation or the flow
equation \cite{polchinski},
\be \label{flow}
\pal\litl[\p]=\left.-\frac12\intdp\pal\dl(p)
       \left[\frac{\d^2\litl}{\d\p(p)\d\p(-p)}
      -\frac{\d\litl}{\d\p(p)}\frac{\d\litl}{\d\p(-p)}\right]
      \right|_{\rm field\mbox{-}dep.\ part}\,.
\ee
To fix appropriate boundary conditions for $\litl$, renormalization
conditions in the conventional renormalization procedure may be used at
$\LLL=0$ where physical Green functions are generated.

Expanding $\litl$ in powers of $\a$, we now define $\liln$'s as the
expansion coefficients, i.e.,
\be
\litl[\p]=\sum_{|N|\ge0}\frac{\a^N}{N!}\liln[\p]\,,
      \quad\left(\lilg{0}[\p]\eq \lil[\p]\right)\,.
\ee
Then, \
$\liln[\p]\;$ is the generating functional of amputated
connected Green functions with an
insertion of an operator $O_N$ which is
identified as the $N$-th order coefficient of $e^{L_C^\lo[\p]}$, i.e.,
\be \label{oiexpansion}
\sum_{|N|\ge1}\frac{\alpha^N}{N!}O_N\eq e^{L_C^\lo}
 =\exp{\bigl(\sum_{|N|\ge1}\frac{\alpha^N}{N!}\lilon\bigr)}\,.
\ee
Explictly, $O_N$ can be calculated as
\be\label{oi}
O_N=\sum_{\sum_{I_l} I_ln_I=N_l}N!\prod_{|I|\ge1}\frac{1}{n_I!}
  \left(\frac{\lilog{I}}{I!}\right)^{n_I}\,,
\ee
where $I=(I_1,\ldots,I_k)$ and $I!=I_1!\cdots I_k!$.
For example, if $k=1$, $O_1=\lilog{1}$, $O_2=\lilog{2}+(\lilog{1})^2$,
$O_3=\lilog{3}+3\lilog{2}\lilog{1}+(\lilog{1})^3$ and so on.
At the zeroth order of $\a$, we obtain the flow equation for the effective
Lagrangian $\lil$ from Eq.~(\ref{flow}), which assumes an identical form as
Eq.~(\ref{flow}) other than the replacement $\litl\ra\lil$.
At order $\a^N$, on the other hand, we have
\bea \label{flowk}
\pal\liln
 &=&-\frac12\intdp\pal\dl(p)\left[\frac{\d^2\liln}{\d\p(p)\d\p(-p)}
    -2\frac{\d\lil}{\d\p(p)}\frac{\d\liln}{\d\p(-p)}\right.\no\\
    &&\hspace{50mm}\left.\left.
    -\sum_{0<I<N}\pmatrix{N\cr I}\frac{\d L^{\LLL;I}}{\d\p(p)}
     \frac{\d L^{\LLL;N-I}}{\d\p(-p)} \right]
     \right|_{\rm field\mbox{-}dep.\ part}\,,
\eea
where $\pmatrix{N\cr I}\eq\pmatrix{N_1\cr
I_1}\cdots\pmatrix{N_k\cr I_k}$ and $I\le N$ means $I_i\le N_i$
(for every $i=1,\ldots,k$) while $I<N$ represents $I\le N$ and $|I|<|N|$.
If $k= |N|=1$, i.e., for a single insertion of operator through
$L^{\lo;1}$, the last term
vanishes and so we have a linear and homogeneous equation
in $L^{\LLL;1}$,
which has been used in discussing
composite operators \cite{hughes,keller}.
Equation (\ref{flowk}) with $k=2$ and $N=(1,1)$ has
also appeared in the literature \cite{keller} in deriving the
short-distance expansion of two composite operators. Note that, if $|N|>1$,
Eq.\ (\ref{flowk}) contains inhomogeneous terms.
However, in spite of the inhomogeneous piece,
it is still of first order in $\liln$ while the homogeneous part
remains
exactly the same for all $N$. Therefore the general solution will be a
sum of a particular solution and a solution to the homogeneous equation (which
is just the $|N|=1$ equation for Green functions with a single insertion
of an operator). This property will be used in the Appendix where
Zimmermann's
normal product and multiple insertion of local vertices are discussed in this
formalism.
Equation (\ref{flowk}) is the starting point of our subsequent analyses.
\setcounter{equation}{0}
\section{STRUCTURES OF THE FULL THEORY}
\subsection{Defining the full theory}
\ \indent As our full theory we consider a scalar theory ($\p$-$\ps$ theory)
which involves two real scalar fields $\p$ and $\ps$,
defined with an ultraviolet cutoff $\lo$.
The mass $M$ of $\ps$ field is assumed to be much larger
than the mass $m$ of $\p$ field but of course
much smaller than
$\lo$.
In the next section, we will discuss the effect of the heavy field $\ps$
at low energy.
The bare action for the system reads
\be
S^{(f)\lo}=\frac12\lbra\p,\dmlo^{-1}\p\rket+\frac12\lbra\ps,\dMlo^{-1}\ps\rket
      +\lfilo\,,
\ee
where $\dmlo$ and $\dMlo$ are respective free-particle cutoff propagators.
We assume that particles interact via quartic couplings, observing $Z_2\times
Z_2$ symmetry. Explicitly, we may write
\be \label{sbare}
\lfilo=\intdx\left[\frac{\r_1^f}2\p^2+\frac{\r_2^f}2(\pa_\m\p)^2
     +\frac{\r_3^f}{4!}\p^4+\frac{\r_4^f}2\ps^2
  +\frac{\r_5^f}2(\pa_\m\ps)^2
     +\frac{\r_6^f}{4!}\ps^4+\frac{\r_7^f}4\p^2\ps^2\right]\,.
\ee
The superscript $f$ is used for quantities of the full theory.
As is conventional, the bare couplings
$\r_a^f$ ($a=1,2,\cdots,7)$ may be
written as
\be \label{barec}
\begin{array}{lll}
\r_1^f=Z_\p m_0^2-m^2,&\r_2^f=Z_\p-1,&\r_3^f=Z_\p^2 g_1^0,\\[2mm]
\r_4^f=Z_\ps M_0^2-M^2,&\r_5^f=Z_\ps-1,&\r_6^f=Z_\ps^2 g_2^0,\\[2mm]
\r_7^f=Z_\p Z_\ps g_3^0\,,&&
\end{array}
\ee
where $(m_0^2,M_0^2)$, $(Z_\p,Z_\ps)$ and $(g_1^0,g_2^0,g_3^0)$ are bare
masses, wave-function renormalizations and bare coupling constants,
respectively. Propagators $\dmlo$ and $\dMlo$ are defined by
\be
\dmlo=\frac{R_m(\lo,p)}{p^2+m^2},\quad
\dMlo=\frac{R_M(\lo,p)}{p^2+M^2}\,,
\ee
where $R_m(\LLL,p)$ and $R_M(\LLL,p)$ are smoothed variants of the sharp cutoff
function $\theta(\LLL-p)$. As
indicated in Sec.\ 2, choosing $R_m$ and
$R_M$ corresponds to choosing a ``renormalization scheme'' and physical
quantities are not affected. Here we
choose the following ``mass-dependent scheme'':
\bea \label{cutoff}
R_m(\LLL,p)&=&\left[1-K\left((1+\LLL/m)^2\right)\right]
               K\left(\frac{p^2}{\LLL^2}\right)\,,\no\\
R_M(\LLL,p)&=&\left[1-K\left(\LLL^2/M^2\right)\right]
               K\left(\frac{p^2}{\LLL^2}\right)\,,
\eea
where
$K$ is a $C^\infty$-function on $[0,\infty)$ such that $K(a)=1$, $0\le a\le1$
and $K(a)=0$, $a\ge4$ with $K(a)\ra1(0)$ exponentially as $a\ra1(4)$.
Thus, both $R_m(\LLL,p)$ and $R_M(\LLL,p)$ are $C^\infty$-functions on
$([0,\infty)\times \bbR^4)$ and $R_m(R_M)$
approaches 0 exponentially as $\LLL\ra0$ (as $\LLL\ra M$).
If $\LLL>m$ for $R_m$ or $\LLL>2M$ for
$R_M$, they simply reduce to $K(p^2/\LLL^2)$.
A notable property of $R_M$ is that if $\LLL<M$, we have $R_M=0$ identically.
This choice of $R_M$ is very natural for our purpose, because it implies
that at the scale $\LLL=M$ we have all modes of the heavy field
$\ps$ integrated out and there remains only the light field $\p$ below
the scale $M$; it
explicitly implements Wilson's point of view on effective field theory.
Therefore, if $\LLL<M$, the flow equation
will look like that for the theory having the
light
field only.
One may regard this property specific to our choice
(\ref{cutoff}) as the
analogy of the so-called ``manifest decoupling'' in
the conventional approach \cite{collins}. On the contrary, if one chose a
``mass-independent scheme'', i.e. if $R_m$ and $R_M$ were chosen independently
of their masses $m$ and $M$, a substantial part of heavy particle modes would
not be integrated out even below the scale $M$, thus making
subsequent discussions rather complicated.

Now, we define $\lfil[\p,\ps]$ following the general line discussed in
Sec.\ 2. It will then satisfy
\be \label{fflow}
\pal\lfil=-\frac12\intdp\left\{\pal\dml(p)
       \left[\frac{\d^2\lfil}{\d\p(p)\d\p(-p)}
      -\frac{\d\lfil}{\d\p(p)}\frac{\d\lfil}{\d\p(-p)}\right]
+(m\ra M,\ \p\ra\ps)\Biggr\}
      \right|_{\rm field\mbox{-}dep.\ part}.
\ee
We expand $\lfil$ in powers of $\p$ and $\ps$ in momentum space,
utilizing $Z_2\times Z_2$ symmetry,
\bea\label{expansion}
\lfil[\p,\ps]=\sum_{|r|\ge0}\sum_{|n|\ge1}g^r
     \int&&\hspace{-5mm}\prod_{j=1}^{2n_1}\frac{d^4p_j}{(2\pi)^4}
      \prod_{j=1}^{2n_2-1}\frac{d^4p'_j}{(2\pi)^4}
      \p(p_1)\cdots\p(p_{2n_1})\ps(p'_1)\cdots\ps(p'_{2n_2-1})\no\\
    &&\times\lfl{r}{2n}(p_1,\ldots,p_{2n_1};p'_1,\ldots,p'_{2n_2-1})\,,
\eea
where $p'_{2n_2}\eq-\sum_{j=1}^{2n_1} p_j-\sum_{j=1}^{2n_2-1}p'_j$.
Some explanations on our notations are in order: $g\eq(g_1,g_2,g_3)$ are
perturbation-expansion parameters which may be identified as renormalized
coupling constants (see below), $r$ and $n$ represent $(r_1,r_2,r_3)$
and $(n_1,n_2)$, respectively, with $|n|\eq n_1+n_2$, $|r|\eq r_1+r_2+r_3$
and
finally, $g^r\eq g_1^{r_1}g_2^{r_2}g_3^{r_3}$.
Then
the vertex function $\lfl{r}{2n}$ has Bose symmetry and, in addition, will
possess following properties:
(i) as we noted in Sec.\ 2, $\lfo{2}{2n}$ can be identified with
amputated connected Green functions $G_{r,2n}^{(f)c}$ of the theory, i.e.,
\be
\lfo{r}{2n}(p_1,\ldots,p_{2n_1};p'_1,\ldots,p'_{2n_2-1})
 =G_{r,2n}^{(f)c}(p_1,\ldots,p_{2n_1};p'_1,\ldots,p'_{2n_2-1})\,;
\ee
(ii) $\lfl{r}{2n}=0$ if $|n|>|r|+1$
since just the connected
diagrams contribute to $\lfl{r}{2n}$ as we have seen in Sec.\ 2;
(iii) $\lfl{r}{2n}\in C^\infty([0,\lo]\times\bbR^{4(n-1)})$ and is
invariant under the $O(4)$ rotation group.
For later use, we also denote vertex functions obtained by summing over
$r_2$ and $r_3$
as $\bar{\cal L}$, i.e.,
\be\label{lbl}
\lbl{r_1}{2n}\eq\sum_{r_2,r_3}g_2^{r_2}g_3^{r_3}\lfl{(r_1,r_2,r_3)}{2n}\,.
\ee

We insert the expansion (\ref{expansion}) into the flow equation (\ref{fflow})
and arrive at the equation
\bea \label{lflflow}
\lefteqn{\pal\lfl{r}{2n}(p_1,\ldots,p_{2n_1};p'_1,\ldots,p'_{2n_2-1})}\no\\
&&=-\pmatrix{2n_1+2\cr 2}\intdq \pal\dml(q)
 \lfl{r}{2(n_1+1,n_2)}(p_1,\ldots,p_{2n_1},q,-q;p'_1,\ldots,p'_{2n_2-1})\no\\
&&\ +2\sum_{r'+r''=r\atop {l_1+l'_1=n_1+1\atop l_2+l'_2=n_2}}l_1l'_1\pal\dml(P)
 \left[\lfl{r'}{2(l_1,l_2)}(p_1,\ldots,p_{2l_1-1},P;p'_1,\ldots,p'_{2l_2-1})
\right.\no\\[-7mm]
&&\left.\hspace{47mm}\times
       \lfl{r''}{2(l'_1,l'_2)}(p_{2l_1},\ldots,p_{2n_1},P;
                     p'_{2l_2+1},\ldots,p'_{2n_2-1})\right]_{symm} \no\\
&&\ -\pmatrix{2n_2+2\cr 2}\intdq \pal\dMl(q)
 \lfl{r}{2(n_1,n_2+1)}(p_1,\ldots,p_{2n_1};p'_1,\ldots,p'_{2n_2},q)\no\\
&&\
 +2\sum_{r'+r''=r\atop {l_1+l'_1=n_1\atop l_2+l'_2=n_2+1}}l_2l'_2\pal\dMl(P')
 \left[\lfl{r'}{2(l_1,l_2)}(p_1,\ldots,p_{2l};p'_1,\ldots,p'_{2l_2-1})
\right.\no\\[-7mm]
&&\left.\hspace{49mm}\times
       \lfl{r''}{2(l'_1,l'_2)}(p_{2l_1+1},\ldots,p_{2n_1};
                     p'_{2l_2},\ldots,p'_{2n_2})\right]_{symm}\,,
\eea
where $P=-\sum_{j=1}^{2l_1-1}p_j-\sum_{j=1}^{2l_2}p'_j$,
$P'=-\sum_{j=1}^{2l_1}p_j-\sum_{j=1}^{2l_2-1}p'_j$ and $[\cdots]_{symm}$
implies symmetrization with respect to momenta $p_1,\ldots,p_{2n_1}$ (and
momenta $p'_1,\ldots,p'_{2n_2}$). Denoting $\p$($\ps$)-legs by
thin(thick)-lines and $\pal\dl$ by a straight line, we can represent this
equation by the diagram shown in Fig.\ 1.
Here some remarks
on Eq.\ (\ref{lflflow})
are in order. First, since we have $\lfl{0}{2n}=0$ (see below)
the restriction $r'+r''=r$ also implies that $r',r''<r$ (i.e.,
$|r'|,|r''|<|r|$ and $r'_i,r''_i\le r_i$).
Also, in relation with the
effective
theory of the next section, it should be noted that,
for $\lfl{r}{2(n_1,0)}$ (i.e., no external heavy particles),
the last term
is identically zero and
$\ps$ enters
the flow equation only through the $\LLL$-differentiated propagator
in the third term.

If suitable boundary conditions are given, the flow equation (\ref{lflflow})
will produce the amputated connected Green functions of the theory,
$G_{r,2n}^{(f)c}=\lfo{r}{2n}$.
{}From Eq.~(\ref{sbare}), we have
\bea \label{fbc}
\LLL=\lo:\hspace{8mm}\pa_p^w\lflo{r}{2n}=0,\hspace{50mm}&&2|n|+|w|>4\,,\no\\
\pa_p^w\lflo{r}{2n}(0)\sim
 \mbox{bare couplings in Eq.~(\ref{barec})},&&2|n|+|w|\le 4\,,\\
 \mbox{(to be determined)}&&\no
\eea
where $w$ is the multiindex
$\{w\}\equiv\{w_1,\ldots,w_{2n_1},w'_1,\ldots,w'_{2n_2-1}\}$,
$w_j=(w_{j1},\ldots,w_{j4})$
and $\pa_p^w\eq\pa_{p_1}^{w_1}\cdots\pa_{p_{2n_1}}^{w_{2n_1}}
     \pa_{p'_1}^{w'_1}\cdots\pa_{p'_{2n_1}}^{w'_{2n_2}}$,
$\pa_{p_i}^{w_i}\equiv\prod_{\m=1}^4\pa^{w_{i\m}}/(\pa p_{i\m})^{w_{i\m}}$.
We will also
write $|w|\equiv\sum_{i,\m}(w_{i\m}+
w'_{i\m})$.
At $\LLL=0$, on the
other hand, we impose following renormalization conditions on relevant
terms\footnote{With these
conditions the expansion parameters $(g_1,g_2,g_3)$ are now identified with
renormalized coupling constants. As far as our proof goes, details of
the renormalization conditions are not important as long as one choose the
renormalization points for the light particle at momenta much smaller than $M$
\cite{collins}. But, for practical applications, it would be necessary
to choose the renormalization conditions for the heavy particles at momentum
values of order the heavy particle mass \cite{ovrut}.}:
\be \label{fbc2}
\begin{array}{lll}
\lfo{r}{(2,0)}(0)=0,&
 \pa_{p_\m}\pa_{p_\n}\lfo{r}{(2,0)}(0)=0,&\hspace{3mm}
 \lfo{r}{(4,0)}(0)=\frac1{4!}\d_{r,(1,0,0)}\,,\\[3mm]
\lfo{r}{(0,2)}(\bar p'_1)=0,
 & \pa_{p_\m}\pa_{p_\n}\lfo{r}{(0,2)}(\bar p'_1)=0,&\hspace{3mm}
 \lfo{r}{(0,4)}(\bar p'_2,\bar p'_3,\bar p'_4)
                        =\frac1{4!}\d_{r,(0,1,0)}\,,\\[3mm]
\lfo{r}{(2,2)}(0,\bar p'_5)=\frac14\d_{r,(0,0,1)}.&&
\end{array}
\ee
Here, normalization momenta ${\bar p'}_i$ ($i=1,\ldots,5$) are chosen to  be
constants of magnitude $M$; i.e.,
we have chosen the renormalization points for the light-particle Green
functions at zero momentum, and those for the heavy particles at momenta of
magnitude $M$. It is then
easy to show that
$\r_a^f$ are uniquely fixed
within perturbation theory.
(See, for example, Ref.\ \cite{keller}.)

\subsection{Bounds on Green functions}
\ \indent
Now the theory being completely specified, one may
use the flow equation to demonstrate the perturbative renormalizability
\cite{polchinski,keller}, which amounts to boundedness and convergence
of the vertex functions $\lfl{r}{2n}$
as $\lo\ra\infty$. To reach our goal in the next
section, however, we need slightly stronger bounds which
take particle-mass dependence into account explicitly.

First, we introduce a set of norms on vertex functions $\|\ \|_{(a,b,b')}$,
for positive real $a$, $b$, and $b'$,
\be \label{norm}
\norm{\pa^{z}\lfl{r}{2(n_1,n_2)}}{(a,b,b')}\eq
 \max_{|p_i|\le \max{\{a,b\}}\atop {|p'_i|\le\max\{a,b'\}\atop{|w|=z}}}
 |\pa_p^w\lfl{r}{2(n_1,n_2)}(p_1,\ldots,p_{2n_1};p'_1,\ldots,p'_{2n_2-1})|\,.
\ee
For vertices with no heavy-particle leg, we will often omit $b'$,
$ \norm{\pa^z\lfl{r}{2(n_1,0)}}{(a,b)}\eq
 \norm{\pa^z\lfl{r}{2(n_1,0)}}{(a,b,b')}.  $
Now it is not difficult to show that, with the choice (\ref{cutoff}), we
have the following uniform bounds:
\be \label{dbound}
\left|\frac{\pa_p^z\pal R_m(\LLL,p)}{(p^2+m^2)^n}\right|\le
 \frac{c}{(\LLL+m)^{2n+z+1}}\,,\qquad
\left|\frac{\pa_p^z\pal R_M(\LLL,p)}{(p^2+M^2)^n}\right|\le
        \frac{c'}{\LLL^{2n+z+1}}\theta(\LLL-M)\,,
\ee
where $0\le\LLL\le\lo$ and $c$, $c'$ are constants which do not depend on
$\LLL,\lo,\ M,$ or $m$
(but may depend on $z,\ n$ and $k$).
Acting $\pa_p^w$ on Eq.\ (\ref{lflflow})
and estimating the resulting expression with the help of the
bounds (\ref{dbound}), we find that, for a fixed constant $\m$ of order $m$,
\bea \label{flbh}
\normn{\pal\pa^z\lfl{r}{2n}}&&\\
 &&\hspace{-38mm}\le \mbox{const}\Biggl\{(\LLL+m)
\left[\normn{\pa^z\lfl{r}{2(n_1+1,n_2)}}+
    \theta(\LLL-M)\normn{\pa^z\lfl{r}{2(n_1,n_2+1)}}\right]\no\\
 &&\hspace{-38mm}+\Biggl(
   \sum_{r'+r''=r \atop {l+l'=(n_1+1,n_2) \atop z_1+z_2+z_3=z}}\hspace{-4mm}
  +\ \ \theta(\LLL-M)\hspace{-7mm}
    \sum_{r'+r''=r \atop {l+l'=(n_1,n_2+1) \atop z_1+z_2+z_3=z}}
  \Biggr)
   \frac{1}{(\LLL+m)^{3+z_1}}\normn{\pa^{z_2}\lfl{r'}{2(l_1,l_2)}}
        \normn{\pa^{z_3}\lfl{r''}{2(l'_1,l'_2)}}\Biggr\}\,,\no
\eea
where ``const'' stands for some finite number
independent of
$M,\ \LLL$ and $\lo$ (but may
depend on $m$ and $\m$ through
$(\m/m)$). Equation (\ref{flbh})
holds for all $\LLL\in[0,\lo]$, and
it allows us to present our
subsequent results and proofs in simpler forms than those appeared in
earlier works \cite{polchinski,keller}.

Now we are ready to prove the boundeness of vertex functions:
\begin{theorem}\label{theorem1}
For vertices with no external heavy-particle leg,
\be \label{11a}
\normn{\pa^z\lfl{r}{2(n_1,0)}}
     \le(\LLL+m)^{4-2n_1-z}\plog{\LLL+m}{m}\,,\quad 0\le\LLL\le\lo\,,
\ee
while, for irrelevant vertices ($2|n|+z>4$) with external heavy-particle legs,
\be \label{11b}
\normn{\pa^z\lfl{r}{2n}}
     \le(\LLL+m)^{4-2|n|-z}\plog{\LLL+M}{m}\,,\quad 0\le\LLL\le\lo\,,
\ee
and, finally, for relevant vertices ($2|n|+4\le4$) with external
heavy-particle legs,
\be \label{11c}
\normn{\pa^z\lfl{r}{2n}}
     \le(\LLL+M)^{4-2|n|-z}\plog{\LLL+M}{m}\,,\quad 0\le\LLL\le\lo\,,
\ee
where $\mbox{\rm Plog}(\cdot)$ is some polynomial in $\log(\cdot)$ whose
coefficients are independent of $M$, $\LLL$, and $\lo$.
\end{theorem}
Before proving Theorem \ref{theorem1}, we here note that the above bounds
are not unexpected --- they just show the right scaling behaviors. That is,
the effect of imposing renormalization conditions at momenta of order
$M$ for heavy-particle legs
shows with appropriate powers of $M$ (up to logarithmic corrections) in the
flow
equation of vertices with external heavy-particle legs.
On the other hand, in the light-particle sector, the
bounds have the same form as those of a single scalar theory
\cite{polchinski,keller}: the large $M$ corrections do not show up
because
$m$ is forced to be small by hand.
Of course, this should be the case for decoupling to occur in the first
place.

{\it Proof of Theorem \ref{theorem1}.\/}
We proceed along the induction scheme set up in Ref.\
\cite{polchinski}. Equations (\ref{11a})--(\ref{11c}) are trivially true
for $r=0$ or for $(r,n)$ with $|r|>0$ and $|n|>|r|+1$,
because $\lfl{r}{2n}=0$ for those $(r,n)$-values as we have seen before. Now
suppose it holds for $(r,n)$ satisfying the condition
$|r|\le s-1$, or $|r|=s$ and $|n|\ge k+1$. Then the induction step consists
in proving Eqs.~(\ref{11a})--(\ref{11c}) for
any fixed
$(r,n)$ with $|r|=s$ and $|n|=k$.
We begin with the irrelevant vertices, i.e., with the case $2k+z>4$.
For these we
use the boundary condition (\ref{fbc}) to write
\be \label{pf11}
\normn{\pa^z\lfl{r}{2n}}
 \le\int_\LLL^\lo d\lll \normn{\pa_\lll\pa^z\lfll{r}{2n}}\,.
\ee
The right hand side is now estimated with the help of Eq.~(\ref{flbh}).
If $\LLL>M$, according to the induction hypothesis it is easy to
find that, for irrelevant vertices with no heavy-particle leg\footnote{Each
time the symbol Plog appears, it stands in general for a new
polynomial of log.},
\bea \label{pf12}
\normn{\pa^z\lfl{r}{2(n_1,0)}}
 &\le&\int_\LLL^\lo d\lll (\lll+m)^{3-2k-z}\plog{\lll}{m}\no\\
 &\le&(\LLL+m)^{4-2k-z}\plog{\LLL+m}{m}\,.
\eea
A similar estimate yields the bound (\ref{11b}) for irrelevant vertices with
heavy-particle legs. For $\LLL\le M$, the bounds on $\pa_p^w\lfl{r}{(0,2)}$
with $z=0,1$ in Eq.\ (\ref{11c}) look problematic because they are not
suppressed by appropriate powers of $\LLL$, i.e., they might result in
undesirable extra powers of $M$ when one estimates the right hand side of
Eq.\ (\ref{flbh}) with the induction hypothesis. They, however, cause no real
problem since
they appear in the right hand side of Eq.\ (\ref{flbh}) only with
$\theta(\LLL-M)$. There are other dangerous terms when one estimates vertices
with $n_2=0$ for $\LLL<M$, namely, the $\log M/m$ corrections in Eqs.\
(\ref{11b}) and (\ref{11c}); they are again multiplied by $\theta(\LLL-M)$
and do not contribute at all if $\LLL<M$.
Therefore,
we conclude that the induction argument used in $\LLL> M$ case  goes through
with no obstuction for irrelevant vertices.

For the case of relevant vertices ($2k+z\le4$), we begin with
$\lfl{r}{(0,4)}$.
Recalling
the boundary conditions at $\LLL=0$, we
integrate the flow equation (\ref{lflflow})
from 0 to $\LLL$ (at
$p'_i={\bar p}'_i$) to obtain
\bea
|\lfl{r}{(0,4)}({\bar p}'_i)|&\le&
     |\lfo{r}{(0,4)}({\bar p}'_i)|+\int_0^\LLL
                 d\lll\normn{\pa_\lll\lfll{r}{(0,4)}}\no\\
 &\le& \plog{\LLL+M}{m}\,,
\eea
where we have used the induction hypothesis in the second line.
Now, using the Taylor's formula
\be \label{taylor}
\lfl{r}{(0,4)}(p'_i)=\lfl{r}{(0,4)}({\bar p}'_i)
   +\sum_{i=1}^3(p'_i-{\bar p}'_i)
         \int_0^1 dt\pa_{q_{i\m}}\lfl{r}{(0,4)}(q_i)|_{q_i=tp_i}\,,
\ee
and then putting $\LLL=M$,we get a bound for ${\cal L}_{r,(0,4)}^{(f)\LLL=M}$
as being of order $\mbox{Plog}(M/m)$ at most. Then the integration of
$\pa_\lll\lfll{r}{(0,4)}$ from $\LLL$ to $M$ gives Eq.\ (\ref{11c}). The same
argument works for other relevant vertices with $2k+z=4$.
Next, if $n=(0,1)$ and $z=1$, we may use the property that
$\pa_p^w\lfl{r}{2n}(0)=0$ for odd dimensional vertices (due to $Z_2\times Z_2$
symmetry), and therefore
\bea \label{oddlfl}
\normn{\pa\lfl{r}{(0,2)}}&=&
 \normn{p'_\m\int_0^1 dt
        \pa_{q'_\m}\pa_{q'}\lfl{r}{(0,2)}(q')|_{q'=tp'}}\no\\
 &\le&(\LLL+M)\plog{\LLL+M}{m}\,.
\eea
The same argument shows that
$\normn{\pa\lfl{r}{(2,0)}}\le(\LLL+m)\plog{\LLL+m}{m}$. It is then
straightforward to obtain the bounds for $|n|=1$ and $z=0$ using Taylor
expansion around the renormalization point, as we did for $2k+z=4$ case.
\qedbox

In particular, at $\LLL=0$, Theorem 1 tells us that amputated connected Green
functions and their derivatives of the theory are bounded,
for $|p_i|\le\m$ and $|p'_i|\le M$, by
\be
|\pa_p^wG_{r,2n}^{(f)c}(p_1,\ldots,p_{2n_1};p'_1,\ldots,p'_{2n_2-1})|
  \le \cases{
     \mbox{const}\cdot m^{4-2n_1-|w|}\,,\hspace{14mm} n_2=0\cr
     m^{4-2|n|-|w|}\plog{M}{m},\qquad n_2\neq0,\ 2|n|+|w|>4\cr
     M^{4-2|n|-|w|}\plog{M}{m},\qquad n_2\neq0,\ 2|n|+|w|\le4.}
\ee

Now we show that Green functions converges to a finite limit as the UV
cutoff $\lo$ goes to infinity. It will also complete the proof of the
perturbative renormalizability of the $\p$-$\ps$ theory.
\begin{theorem}\label{theorem2}
For vertices with no external heavy-particle leg,
\be \label{21a}
\normn{\palo\pa^z\lfl{r}{2(n_1,0)}}
     \le\left(\frac{\LLL+m}{\lo}\right)^3
       (\LLL+m)^{3-2n_1-z}\plog{\lo}{m}\,,\quad 0\le\LLL\le\lo\,,
\ee
while, for irrelevant vertices ($2|n|+z>4$) with external heavy-particle legs,
\be \label{21b}
\normn{\palo\pa^z\lfl{r}{2n}}
     \le\left(\frac{\LLL+M}{\lo}\right)^2\left(\frac{\LLL+m}{\lo}\right)
      (\LLL+m)^{3-2|n|-z}\plog{\lo}{m}\,,\hspace{2mm} 0\le\LLL\le\lo\,,
\ee
and, finally, for relevant vertices ($2|n|+4\le4$) with external
heavy-particle legs,
\be \label{21c}
\normn{\palo\pa^z\lfl{r}{2n}}
     \le\left(\frac{\LLL+M}{\lo}\right)^3
      (\LLL+M)^{3-2|n|-z}\plog{\lo}{m}\,,\quad 0\le\LLL\le\lo\,.
\ee
\end{theorem}

{\it Proof.\/} The method is very similar to that employed in Theorem 1 and we
only sketch the procedure.
For irrelevant vertices, we use the identity
\be
\palo\pa_p^w(\lflo{r}{2n}-\lfl{r}{2n})
 =\pa_p^w\pal\lfl{r}{2n}|_{\LLL=\lo}
     +\int_\LLL^\lo d\lll\pa_\lll\palo\pa_p^w\lfll{r}{2n},
\ee
while, for relevant vertices, we use the relation
\be
\palo\pa_p^w(\lfl{r}{2n}-\lfo{r}{2n})
 =\int_0^\LLL d\lll\pa_\lll\palo\pa_p^w\lfll{r}{2n}\,.
\ee
Then, differentiating Eq.\
(\ref{lflflow}) with respect to
$\lo$, we obtain
the bounds for the right hand sides, as in
Eq.~(\ref{flbh}).
Finally, applying the induction scheme as described in Theorem 1 and using
the results of Theorem 1, it is straightforward to prove the theorem.
\qedbox
\setcounter{equation}{0}
\section{DECOUPLING, FACTORIZATION AND EFFECTIVE LAGRANGIANS}
\ \indent
We are now ready to discuss the low-energy physics of the
$\p$-$\ps$ theory.
In this section we will prove the decoupling theorem and the factorization
of virtual heavy-particle effects to any desired order in $1/M^2$
on the basis of the flow equation
approach. We start with proving the decoupling theorem which corresponds to
the zeroth order result of factorization.
\subsection{Low-energy effective field theory and the decoupling theorem}
\ \indent
Let us consider a theory of light particles (the $\p$ theory) of mass $m$
with the interaction part $\lilo$ of the bare action given by
\be  \label{cebare}
\lilo=\intdx\left[\frac12\r_1\p^2(x)+\frac12\r_2(\pa_\m\p(x))^2
        +\frac{1}{4!}\r_3\p^4(x)\right]\,.
\ee
As before, $\lil[\p]$ is defined to satisfy the flow equation (\ref{flow})
and can be expanded as
\be \label{eexpansion}
\lil[\p]=\sum_{r=0}^\infty\sum_{n=1}^\infty g_1^r
          \int\prod_{j=1}^{2n-1}\frac{d^4p_j}{(2\pi)^4}\p(p_1)\cdots\p(p_{2n})
         \lvl{r}{2n}(p_1,\ldots,p_{2n-1})\,.
\ee
The vertex function $\lvl{r}{2n}$ has properties similar to those for
$\lfl{r}{2n}$: $\lvl{r}{2n}$ has Bose symmetry,
$\lvl{r}{2n}=0$ if $n>r+1$, $\lvl{r}{2n}$ is $C^{\infty}$ and
Euclidean-invariant, and
finally $\lvo{r}{2n}$ is equal to the amputated connected
Green function of the theory, i.e.,
$G_{r,2n}^{c}$. It satisfies the flow equation
\bea\label{lvlflow}
\pal\lvl{r}{2n}(p_1,\ldots,p_{2n-1})
&\hspace{-2.5mm}=&\hspace{-2.5mm}-\pmatrix{2n+2\cr 2}\intdq \pal\dml(q)
 \lvl{r}{2(n+1)}(q,-q,p_1,\ldots,p_{2n-1})\\
& &\hspace{-8mm}+2\sum_{r'+r''=r\atop l+l'=n+1}ll'\pal\dml(P)
 \left[\lvl{r'}{2l}(p_1,\ldots,p_{2l-1})
       \lvl{r''}{2l'}(-P,p_{2l},\ldots,p_{2n-1})\right]_{symm}\;,\no
\eea
where $P=-\sum_{j=1}^{2n-1}p_j$.
The boundary conditions which $\lvl{r}{2n}$'s obey are determined as follows.
First, at $\LLL=\lo$, irrelevant vertices vanish, i.e.,
\be
\pa_p^w\lvlo{r}{2n}=0,\quad 2n+|w|>4\,.
\ee
At $\LLL=0$, we impose the same renormalization conditions (\ref{fbc2})
as those in the full theory,
\be \label{effbc}
\lvo{r}{2}(0)=\pa_{p_\m}\pa_{p_\n}\lvo{r}{2}(0)=0,\qquad
\lvo{r}{4}(0)=\frac1{4!}\d_{r1}\,,
\ee
so that the $\p$ theory may not be distinguishable from the $\p$-$\ps$
theory at low energies.

{}From Theorems \ref{theorem1} and \ref{theorem2} we immediately have
the bounds on the vertex functions
$\lvl{r}{2n}$ (which amounts to proving the
perturbative renormalizabililty of the $\p$ theory), viz.,
\bea \label{th1a}
&&\norml{\pa^z\lvl{r}{2n}}\le(\LLL+m)^{4-2n-z}\plog{\LLL+m}{m}\,,\qquad
                        0\le\LLL\le\lo\,,\\
&&\norml{\pa_\lo\pa^z\lvl{r}{2n}}
     \le\left(\frac{\LLL+m}{\lo}\right)^3(\LLL+m)^{3-2n-z}\plog{\lo}{m}\,,
     \qquad 0\le\LLL\le\lo\,. \label{th2a}
\eea

Now we are in a position to prove the decoupling theorem:
\begin{theorem}\label{theorem3}
The difference of the vertex functions $\lfl{r}{2(n,0)}$ in the
$\p$-$\ps$ theory and $\lvl{r_1}{2n}$ in the
$\p$ theory, where
$r=(r_1,r_2,r_3)$,satisfies the bound
\be\label{th3}
\norml{\pa^z(\lfl{r}{2(n,0)}-\lvl{r_1}{2n}\d_{r,(r_1,0,0)})}
\le\left\{\begin{array}{ll}
        \displaystyle\left(\frac{\LLL+m}{M}\right)^2
                      (\LLL+m)^{4-2n-z}\plog{M}{m}\,,&
                      \hspace{2mm} 0\le\LLL\le M\\[4mm]
        \displaystyle\LLL^{4-2n-z}\plog{\LLL}{m}\,.&
                      M\le\LLL\le\lo\,,\end{array}\right.
\ee
\end{theorem}
{\it Proof.} If $\LLL\in[M,\lo]$, the bound is immediate from Eqs.\
(\ref{11a}) and (\ref{th1a}), and hence it
suffices to consider
$0\le\LLL\le M$ only. Let us
define $\dvl{r}{2n}$ as the difference of vertex
functions,
\be
\dvl{r}{2n}\eq\lfl{r}{2(n,0)}-\lvl{r_1}{2n}\d_{r,(r_1,0,0)}\,.
\ee
Subtracting the flow equation (\ref{lflflow}) for $\lfl{r}{2n}$ from that
for $\lvl{r}{2n}$ (given in (\ref{lvlflow})) and making estimates using Eq.
(\ref{dbound}), one easily finds that, for $0\le\LLL\le M$,
\bea\label{fdlb}
\norml{\pa_\LLL\pa^z\dvl{r}{2n}}
&\le&\mbox{const}\Biggl\{(\LLL+m)\norml{\pa^z\dvl{r}{2(n+1)}}
 \\
 &&\hspace{-35mm}+{\sum_{r'+r''=r\atop{l+l'=n+1\atop z_1+z_2+z_3=z}}}
  \left.\frac{1}{(\LLL+m)^{3+z_1}}
  \norml{\pa^{z_2}\dvl{r'}{2l}}\left(\norml{\pa^{z_3}\dvl{r''}{2l'}}
 +\norml{\pa^{z_3}\lvl{r''_1}{2l'}}\d_{r'',(r''_1,0,0)}\right)\right\}\,,\no
\eea
where it should be noted
that terms involving the heavy-particle propagator all vanish.
Boundary conditions for $\dvl{r}{2n}$ are simply
\bea\label{dbc}
&&\pa_p^w\dvlo{r}{2n}=0,\quad \mbox{for }2n+|w|>4\,,\no\\
&&\dvo{r}{2}(0)=\pa_{p_\m}\pa_{p_\n}\dvo{r}{2}(0)=\dvo{r}{4}(0)=0\,.
\eea

Since $\dvl{r}{2n}=0$ if $r=0$ or $n>|r|+1$ with $|r|>0$,
the theorem holds trivially for these values of $r,n$.
So we will now prove the validity
of Eq.~(\ref{th3}) assuming that it is true for $r$ and $n$ satisfying
$|r|\le s-1$ or $|r|=s$, $n\ge k+1$. Pick any particular $r$
satisfying $|r|=s$. As usual
we first consider the case of irrelevant vertices.
For $2k+z>5$, we may use the same argument leading to
Eq.~(\ref{pf12}) in the proof of Theorem 1 and find that
\bea
\norml{\pa^z\dvl{r}{2k}}
&\le &\norml{\pa^z{\cal D}_{r,2k}^{M}}
   +\int_\LLL^Md\lll\norm{\pa_\lll\pa^z\dvl{r}{2k}}{(2\lll,\m)}\no\\
&\le & M^{4-2k-z}\plog{M}{m}
  +\int_\LLL^Md\lll\frac{(\lll+m)^{5-2k-z}}{M^2}\plog{M}{m}\no\\
&\le &\frac{(\LLL+m)^{6-2k-z}}{M^2}\plog{M}{m}\,,
\eea
where, in the second line, we used the induction hypothesis.
If $2k+z=5$, we may use the property that
$\pa_p^w\dvl{r}{2k}(0)=0$ for odd dimensional vertices and therefore
\be \label{odd}
\norml{\pa^z\dvl{r}{2k}}=
 \norml{\int_0^1 dt\sum_{i=1}^{2n-1}p_{i\m}\pa_{q_{i\m}}
      \pa_q^w\dvl{r}{2k}(q_1,\ldots,q_{2n-1})|_{q_i=tp_i}}
\ee
to obtain the bound (\ref{th3}). The case with relevant vertices (i.e., with
$2k+z\le4$) is quite trivial.
As in the proof of Theorem \ref{theorem1}, we use the boundary conditions
(\ref{dbc}) at $\LLL=0$ and obtain
\be
|\pa_p^w\dvl{r}{2k}(0)|
  \le\int_0^\LLL d\lll\norm{\pa_\lll\pa^z\dvll{r}{2k}}{(2\lll,\m)}\,.
\ee
Then
Taylor's formula (\ref{taylor}), with $\lfl{r}{2n}$ replaced by $\dvl{r}{2n}$
and also using the argument immediately after Eq.~(\ref{taylor}), yields
the desired bound. \qedbox
\vspace{2mm}

Therefore, the difference between
amputated connected Green functions of
the $\p$-$\ps$ theory and those of the $\p$ theory is bounded by
$\left(\frac{m}{M}\right)^2m^{4-2n-z}\plog{M}{m}$ for $|p_i|\le\m$,
which is the standard statement of the decoupling theorem. Also note
that the difference in the vertex functions is no longer small if $\LLL$
becomes comparable to the heavy particle mass $M$,
which implies that
low-energy effective theory is not useful above the heavy particle mass
scale.

In fact,
as we discussed in Sec.\ 1,
the decoupling theorem
may look more or less obvious if one takes the view that only the light field
is subject to direct observation.
One may then
imagine a pure $\p$-field theory with the Lagrangian given from
the Lagrangian of the $\p$-$\ps$ theory
after integrating out the heavy
field, and identify the cutoff $\lo$ with the heavy particle
mass scale $M$.
What we have shown above is
then that
(i) integration of the heavy field and the ``identification'' of the cutoff
$\lo$ with the heavy particle mass scale can be achieved by our
``mass-dependent scheme'', Eq. (\ref{cutoff}), and (ii) the resulting nonlocal
irrelevant terms are indeed well-behaved so that decoupling theorem may
follow.

\subsection{Factorization of virtual heavy particle effects}
\ \indent
We now examine higher order terms in $1/M^2$, which originate from virtual
heavy-particle interactions.
Let us suppose that we want to describe low-energy physics in the full theory
accurately up to order $1/M^{2N_0}$. ($N_0$ is some positive integer.)
We will show below that we can then factorize all heavy-particle effects to
the given order by making use of
Eq.\ (\ref{flowk}) (with $k=1$) and by appropriately choosing the
operators $\lilon$, $N=1,2,\ldots,N_0$.

Here, $\lilon$'s may be assumed to have the form
\be \label{lilon}
\lilon=\intdx
      \pmatrix{\mbox{local, Euclidean invariant, even polynomials}\cr
      \mbox{in $\p$ and its derivatives, of dim.$\le 4+2N$}}
\,,\qquad N=1,2,\ldots,N_0\,.
\ee
The coefficient of the polynomials will be chosen later so that $\lilon$
carries information appropriate to $O(1/M^{2N})$-effects from the full theory.
Then the sum of $\lilon$'s, viz.,
\be \label{ltilon}
\ltilog{N_0}\eq\lilo+\sum_{N=1}^{N_0}\frac{1}{N!}\lilon
   \ \Bigl(=\ltilog{N_0-1}+\frac{1}{N_0!}\lilog{N_0}\Bigr)\,,
\ee
will reproduce the predictions of the full theory at low energy with an
accuracy of order $1/M^{2N_0}$.
The quantity  $\liln$, which satisfies the flow equation
(\ref{flowk}), may be expanded in powers of $\p$:
\be \label{lilnexpansion}
\liln[\p]=\sum_{r=0}^\infty\sum_{n=1}^\infty g_1^r
         \int\prod_{j=1}^{2n-1}\frac{d^4p_j}{(2\pi)^4}\p(p_1)\cdots \p(p_{2n})
       \lvln{r}{2n}(p_1,\ldots,p_{2n-1})\,.
\ee
We may also define $\ltln{r}{2n}$'s through the same kind of relation.
Then the vertex function
$\lvln{r}{2n}$ is $C^\infty$ and satisfies the properties similar to
those of $\lfl{r}{2n}$;
in particular, there is a finite number $n_0(r,N)$  such that
$\lvln{r}{2n}=0$ if $n>n_0(r,N)$.
$\lvln{r}{2n}$'s obey the flow equation
\bea\label{lvlnflow}
\lefteqn{\pal\lvln{r}{2n}(p_1,\ldots,p_{2n-1})}\no\\
&&=-\pmatrix{2n+2\cr 2}\intdq \pal\dml(q)
 \lvln{r}{2(n+1)}(q,-q,p_1,\ldots,p_{2n-1})\no\\
&&\ +2\sum_{I=0}^N\sum_{r'+r''=r\atop l+l'=n+1}ll'\pal\dml(P)
 \pmatrix{N\cr I}\left[\lvli{r'}{2l}(p_1,\ldots,p_{2l-1})
       \lvlg{N-I}{r''}{2l'}(-P,p_{2l},\ldots,p_{2n-1})\right]_{symm}\;.\no\\
 &&\eq\fvln{r}{2n}(p_1,\ldots,p_{2n-1})\;,
\eea
where $P=-\sum_{j=1}^{2l-1}p_j$.
To complete the definition of $\lilon$, we must also state the boundary
conditions for $\lvln{r}{2n}$. At $\LLL=\lo$, Eq.~(\ref{lilon}) implies that
\be \label{lvlnbc1}
 \pa_p^w\lvlon{r}{2n}=0,\quad \mbox{for }2n+|w|>4+2N\,.
\ee
For $2n+|w|\le4+2N$, we impose the following conditions, recursively in $N$:
\be \label{lvlnbc2}
 \pa_p^w\lvon{r}{2n}(0)=N!\pa_p^w(\lbo{r}{2n}(0)-\ltog{N-1}{r}{2n}(0))\,,
\ee
where $\bar\lag$ is defined in Eq.\ (\ref{lbl}). (Consequently,
$\pa_p^w\lvon{r}{2n}(0)=0$ for the case $2n+|w|<4+2N$.)
It remains to show that the boundary conditions (\ref{lvlnbc1}) and
(\ref{lvlnbc2}) define $\lilon$'s uniquely.
For this, we use Eq.~(\ref{lvlnflow}) together with the conditions
(\ref{lvlnbc1}) and (\ref{lvlnbc2}) to rewrite
\be\label{lvlnbc}
\lvlon{r}{2n}
 =\taylor\lvon{r}{2n}
  +\int_0^\lo d\lll\taylor\fvlln{r}{2n}\,,
\ee
where the symbol $\tau^l$ stands for the Taylor operator defined by
\be
\tau^lf(p_1,\ldots,p_{n})\equiv
 \sum_{j=0}^l\frac{1}{j!}
  \left.\left(\frac{d}{dt}\right)^jf(tp_1,\ldots,tp_n)\right|_{t=0}
             \qquad(\mbox{for }l\ge0)
\ee
(and $\tau^lf\eq0$ for $l<0$).
Then using induction on $N$, $r$, and $n$ one may easily confirm that
$\lilon$'s are defined uniquely.

The theory being completely specified, we will now prove the factorization.
To begin with, we
define $\dvln{r}{2n}$, where $r$ is a multiindex $r=(r_1,r_2,r_3)$ from now on,
as
\be \label{dvln}
\dvln{r}{2n}\eq\lfl{r}{2(n,0)}-\ltpln{r}{2n}
            \qquad\left(=\dvlg{N-1}{r}{2n}-\lvpln{r}{2n}/N!\right)\,,
\ee
where $\lvpln{r}{2n}$ (and $\ltpln{r}{2n}$, correspondingly) is defined by the
expansion,
$\lvln{r_1}{2n}=\sum_{r_2,r_3}g_2^{r_2}g_3^{r_3}\lvpln{(r_1,r_2,r_3)}{2n}$.
The flow equation satisfied by $\dvln{r}{2n}$ will then be the difference
between those given in
Eqs.~(\ref{lflflow}) and (\ref{lvlnflow}).
After some algebra, we find that the equation can be written in the form
\bea\label{dvlnflow}
\pal\dvln{r}{2n}
 &=&-\pmatrix{2n+2\cr 2}\intdq
 \left[\pal\dml\dvln{r}{2(n+1)}
   +\pal\dMl\lfl{r}{(2n,2)}\right]\\
&&+2\sum_{r'+r''=r\atop l+l'=n+1}ll'\pal\dml
 \left[\sum_{I=1}^N\frac{1}{I!}\lvpli{r'}{2l}
          \dvlg{N-I}{r''}{2l'}
+2\lvpl{r'}{2l}
       \dvln{r''}{2l'}
+\dvl{r'}{2l}
       \dvln{r''}{2l'}\right]_{symm}\,,\no
\eea
where, for simplicity, we have omitted the arguments of vertex functions which
are the same as those in Eq.\ (\ref{lflflow}) or (\ref{lvlnflow}).
The boundary conditions satisfied by $\dvln{r}{2n}$ are very simple:
\bea \label{dvlnbc}
\mbox{at }\LLL=\lo:\qquad&\pa_p^w\dvlon{r}{2n}=0\,,
                      \qquad &\mbox{for }2n+|w|>4+2N\,,\no\\
\mbox{at }\LLL=0:\qquad&\pa_p^w\dvon{r}{2n}(0)=0\,,
                      \qquad &\mbox{for }2n+|w|\le4+2N\,.
\eea
Note that if it were not for the inhomogeneous term containing $\lfl{r}{2n}$ in
Eq.~(\ref{dvlnflow}), our boundary condition (\ref{dvlnbc})
would imply that $\dvln{r}{2n}$ should vanish identically. In other words,
a nonzero result for the difference between
the vertex functions of the $\p$-$\ps$ theory and those of the $\p$ theory
is due to the inhomogeneous term which is in fact nonzero only for large
$\LLL$, i.e., for $\LLL\ge M$. As was done in previous sections, we can
obtain an estimate for $\dvln{r}{2n}$:
\bea\label{fdlnb}
\norml{\pa_\LLL\pa^z\dvln{r}{2n}}
 &\hspace{-2mm}\le\hspace{-2mm}&
     \mbox{const}\left\{(\LLL+m)\norml{\pa^z\dvln{r}{2(n+1)}}
   +\theta(\LLL-M)\LLL\normn{\pa^z\lfl{r}{(2n,2)}}\right.\no\\
 &&\hspace{7mm}+{\sum_{r'+r''=r \atop{l+l'=n+1 \atop z_1+z_2+z_3=z}}}
   \frac{1}{(\LLL+m)^{3+z_1}}\left(\sum_{I=0}^N\norml{\pa^{z_2}\dvli{r'}{2l}}
             \norml{\pa^{z_3}\lvplg{N-I}{r''}{2l'}}\right.\no\\[-3mm]
 &&\hspace{41mm}\left.\left.+\norml{\pa^{z_2}\dvln{r'}{2l}}
              \norml{\pa^{z_3}\dvl{r''}{2l'}}\right)\right\}\,.
\eea

Now we state the factorization
\begin{theorem} \label{theorem4}
The difference between two  vertex functions, $\lfl{r}{2(n,0)}$ of the
$\p$-$\ps$ theory and $\ltpln{r}{2n}$ of the $\p$ theory, satisfies the bound
\be\label{th4}
\norml{\pa^z(\lfl{r}{2(n,0)}-\ltpln{r}{2n})}
\le\left\{\begin{array}{ll}
          \displaystyle\left(\frac{\LLL+m}{M}\right)^{2N+2}
                       (\LLL+m)^{4-2n-z}\plog{M}{m}\,,&
                        \hspace{2mm} 0\le\LLL\le M\\[4mm]
          \displaystyle\left(\frac{\LLL}{M}\right)^{2N}
                        \LLL^{4-2n-z}\plog{\LLL}{m}\,,&
                        M\le\LLL\le\lo\,.\end{array}\right.
\ee
\end{theorem}

{\it Proof.} Here we use induction on $N$, $r$ and $n$. The $N=0$ case
has been proved already in Theorem \ref{theorem3}.
Therefore, supposing that Eq.~(\ref{th4}) is true
for $N\le N'-1$ with some $N'\ge1$, we may demonstrate the validity of Eq.\
(\ref{th4}) for $N=N'$. Then the rest of the steps are actually very similar
to those used in proving Theorem \ref{theorem1} or Theorem \ref{theorem3}.
If $r=0$ or if $n>n_0(|r|,N')$ for $r$ with $|r|>0$, we know that
$\lfl{r}{2(n,0)}$ and $\ltplg{N'}{r}{2n}$ (and hence $\dvlg{N'}{r}{2n}$ also)
vanish identically and Eq.~(\ref{th4}) holds trivially for this case.
So we prove the validity of Eq.~(\ref{th4}) for $\pa_p^w\dvlg{N'}{r}{2k}$
with $|r|=s$
under the assumption that it has been proved for
$|r|\le s-1$, and for $|r|=s$ and $n\ge k+1$. Under
this induction hypothesis, Eq.~(\ref{fdlnb}) is now simplified as
\be
\norml{\pa_\LLL\pa^z\dvlg{N'}{r}{2k}}
\le\left\{\begin{array}{ll}
          \displaystyle\frac{(\LLL+m)^{5+2N'-2k-z}}{M^{2N'+2}}\plog{M}{m}\,,
                        \qquad & 0\le\LLL\le M\\[4mm]
          \displaystyle\frac{\LLL^{3+2N'-2k-z}}{M^{2N'}}\plog{\LLL}{m}\,,&
                        M\le\LLL\le\lo\,,\end{array}\right.
\ee
where we used Eq.\ (\ref{dvln}) in estimating
$\norml{\pa^{z_3}\lvplg{N-I}{r}{2k}}$.
Now if $2k+z>4+2N'$, we go downward from $\LLL=\lo$ as before and conclude that
\bea \label{pf42}
\norml{\pa^z\dvlg{N'}{r}{2k}}
&\le&\int_\LLL^\lo d\lll\norm{\pa_\lll\pa^z\dvllg{N'}{r}{2k}}{(2\lll,\m)}\no\\
&\le&\left\{\begin{array}{ll}
\displaystyle\frac{(\LLL+m)^{6+2N'-2k-z}}{M^{2N'+2}}\plog{M}{m}\,,
                        \qquad & 0\le\LLL\le M\\[4mm]
          \displaystyle\frac{\LLL^{4+2N'-2k-z}}{M^{2N'}}\plog{\LLL}{m}\,,&
                        M\le\LLL\le\lo\,.\end{array}\right.
\eea
(If $2k+z=5+2N'$, the upper inequality of Eq.\ (\ref{pf42}) is obtained using
$\pa_p^w\dvlg{N'}{r}{2k}(0)=0$, as was done in the proof of Theorem
\ref{theorem3}.) For $2k+z\le 4+2N'$, we have
\be \label{datzero}
|\pa_p^w\dvlg{N'}{r}{2k}(0)|
 \le\int_0^\LLL d\lll\norm{\pa_\lll\pa^z\dvllg{N'}{r}{2k}}{(2\lll,\m)}\,,
\ee
and so Eq.~(\ref{th4}) follows with the left hand side replaced by
$|\pa_p^w\dvlg{N'}{r}{2k}(0)|$. Then, again, Taylor's formula
(\ref{taylor}) (now applied to $\dvlg{N'}{r}{2k}$)
relates the corresponding vertices with those involving twice more
differentiation by momenta. Therefore,
to complete the proof, induction on $z$ should also be used (for fixed $k$). If
$z>4+2N'-2k$, Eq.~(\ref{th4}) holds as we have just shown above. Now assume
that Eq.~(\ref{th4}) is true for all $z>z_0$ for some $z_0>0$. Then, finally,
Taylor's formula and Eq.\ (\ref{datzero}) guarantee the validity of
Eq.~(\ref{th4}) for any $w$ with $|w|=z_0$.
\qedbox
\vspace{3mm}

If we choose $\LLL=0$, in Eq.~(\ref{th4}), it yields the bound for the
difference between connected
amputated Green functions of the $\p$-$\ps$ theory and those of the $\p$
theory with insertions of appropriate operators defined above. Explicitly,
for $|p_i|\le\m$,
\be
|\pa_p^w(G_{r,2(n,0)}^{(f)c}(p_1,\ldots,p_{2n-1})
             -\widetilde G_{r,2n}^{'c;N_0}(p_1,\ldots,p_{2n-1}))|
 \le\left(\frac{m}{M}\right)^{2N_0+2}m^{4-2n-z}\plog{M}{m},
\ee
where
$\widetilde G_{2n}^{c;N_0}
 =\lttog{N_0}{2n}=\sum_{N=0}^{N_0} G_{2n}^{c;N}/N!$. Thus
one may identify the effective theory which is
accurate up to order $1/M^{2N_0}$ as that defined by
$\ltilog{N_0}=\lilo+\sum_{N=1}^{N_0}\lilon/N!$
and the flow equation (\ref{flowk}).
We will discuss more on this effective  action in Sec.\ 4.3.
If $\LLL$ is larger than $M$, Theorem \ref{theorem4} says
that the vertex function $\pa_p^w\ltln{r}{2n}$ is larger than
$\pa_p^w\ltlg{N=0}{r}{2n}$ by a factor $(\LLL/M)^{2N}$. Therefore, in this
scale, increasing $N$ (or adding more irrelevant terms, equivalently) makes
the vertex functions ``unnaturally'' large and the theory becomes ``less
effective''. This is because we have imposed
unnatural renormalization conditions
(\ref{lvlnbc2}) on irrelevant vertices. If we chose natural values as the bare
irrelevant couplings, they would give only $O((m/\lo)^{2N})$ contributions to
Green functions; but, in Eq.\ (\ref{lvlnbc2}), we imposed the
values of order $(m/M)^{2N}$ to $\ltln{r}{2n}$ as the renormalization
conditions, which are natural only if the cutoff is around $M$. This affirms
that the effective theory is useful only below the heavy particle mass
scale $M$.

The convergence bound as $\lo\ra\infty$ shows a similar behavior. For
$\LLL<M$, the difference converges faster by a factor
$\left(\frac{\LLL+m}{M}\right)^{2N}$ than the $N=0$ case, while the convergence
becomes worse if $\LLL>M$. We state the result here
omitting proof:
\be \label{th5}
\norml{\pa_\lo\pa^z(\lfl{r}{2n}-\ltpln{r}{2n})}
 \le\left(\frac{\LLL+m}{\lo}\right)^3\left(\frac{\LLL+m}{M}\right)^{2N}
             (\LLL+m)^{3-2n-z}\plog{\lo}{m}\,.
\ee
\subsection{Irrelevant operators in the effective Lagrangian}
\ \indent
So far we have shown that, at low energy, connected amputated
Green functions of the full theory are reproduced by considering those of
the $\p$ theory plus those with insertion of operators $O_{N}$,
$N=1,\cdots,N_0$, defined in Eq.\ (\ref{oiexpansion}).
Such operator insertions have a simple interpretation if one
uses the normal product notation \cite{zimmermann,keller}
described in the Appendix. Let us begin with
$N_0=1$. Comparing Eq.\ (\ref{lvlnbc2}) with Eq.\ (\ref{lvlnbc2a}), we identify
$\lilog{1}$ as
\be \label{lilo1}
\lilog{1}=\sum_{2n+|w|=6}b_{2n,\{w\}}^{R;1}[\mono{2n}{w}]\,,
\ee
where
$b_{2n,\{w\}}^{R;1}
 =\frac{1}{\caln}\pa_p^w(G_{2(n,0)}^{(f)c}-G_{2n}^c)(0)$, $\caln$ is a
combinatorial factor defined by Eq.\ (\ref{lvlnbc2a}),
and $[\mono{2n}{w}]$ is the normal product of a local vertex
\be \label{mono}
\mono{2n}{w}\eq\intdx \pa_x^{w_1}\p\cdots\pa_x^{w_{2n-1}}\p\p(x)
\ee
(see Eq.\ (\ref{monobc})).
In terms of the local
Euclidean-invariant vertices of dimension six,
the above equation can be cast into a more attractive form given in Ref.\
\cite{clee},
\bea \label{lilo11}
\lilog{1}
 &=&a_1^{(6)}\intdx[\p(\pa^2)^2\p(x)]+a_2^{(6)}\intdx[\p^3\pa^2\p(x)]
        +a_3^{(6)}\intdx[\p^6(x)]\no\\
 &\eq&\sum_{i=1}^3a_i^{(6)}[\O_i^{(6)}]\,,
\eea
with
\bea \label{ais}
a_1^{(6)}&=&\frac{1}{8\cdot4!}(\pa^2)^2(G_{(2,0)}^{(f)c}-G_2^c)(0)\,,
                \no\\
a_2^{(6)}&=&-\frac{1}{8}\pa^2(G_{(4,0)}^{(f)c}-G_4^c)(0)\,,\no\\
a_3^{(6)}&=&(G_{(6,0)}^{(f)c}-G_6^c)(0)\,.
\eea
The $1/M^2$-order effects
are then described by the
insertion of the operator
$O_1=\sum_i a_i^{(6)}[\O_i^{(6)}]$:
\be
G_{2(n,0)}^{(f)c}=G_{2n}^c+\sum_{i=1}^3a_i^{(6)}G_{2n}^{c;1}([\O_i^{(6)}])
  +O\left(\frac{m^{8-2n}}{M^4}\plog{M}{m}\right)\,,
\ee
where $G_{2n}^{c;1}([\O_i^{(6)}])$ denotes the Green function with
$[\O_i^{(6)}]$ inserted.
For general $N_0$, $\lilon$ $(N=1,\ldots,N_0$) can be identified as
\be \label{lilon1}
\lilon=\sum_{2n+|w|=4+2N}b_{2n,\{w\}}^{R;N}[\mono{2n}{w}]+\lilon_{CT}\,,
\ee
where
\be
b_{2n,\{w\}}^{R;N}
 =\frac{N!}{\caln}\sum_{2n+|w|=4+2N}
  \pa_p^w(G_{2(n,0)}^{(f)c}-\widetilde G_{2n}^{c;N-1})(0)\,,
\ee
and $\lilon_{CT}$ denote counterterms which are needed to cancel the new
divergences due to the multiple insertion of $\lilog{I}$'s,
$I=1,\ldots,N-1$.
(See the discussion of the last paragraph in the Appendix for
explicit forms of $\lilon_{CT}$.) Now let $[\O_i^{(4+2N)}]$'s
form a  complete set of mutually independent Euclidean-invariant local
vertices of dimension $(4+2N)$ and let $a_i^{(4+2N)}$'s be appropriate
expansion coefficients as in Eq.\ (\ref{lilo11}), so that we may write
\be \label{lilon2}
\lilon=\sum_ia_i^{(4+2N)}[\O_i^{(4+2N)}]+\lilon_{CT}\,.
\ee
Then the effective action $\ltilog{N_0}$ may be written as
\bea
\ltilog{N_0}&=&\lilo
               +\sum_{N=1}^{N_0}\sum_i\frac{a_i^{(4+2N)}}{N!}[\O_i^{(4+2N)}]
               +\sum_{N=2}^{N_0}\frac{1}{N!}\lilon_{CT}\no\\
    &=&\ltilog{N_0-1}
      +\sum_i\frac{a_i^{(4+2N_0)}}{N_0!}[\O_i^{(4+2N_0)}]
      +\frac{1}{N_0!}\lilog{N_0}_{CT}\,,
\eea
where $a_i^{(4+2N)}$'s are appropriate linear combinations of
$b_{2n,\{w\}}^{R;N}$'s (all of which are $O(1/M^{2N})$). Thus the
$1/M^{2N}$-order information of the full theory is factorized into the
coefficient $a_i^{(4+2N)}$'s of local vertices of dimension $(4+2N)$.
Given this bare Lagrangian, the flow equation
guarantees that all Green functions are finite and accurate up to order
$1/M^{2N}$.

The interpretation of this result is clear from the viewpoint of the
renormalization group
flow in the infinite dimensional space of possible
Lagrangians.
As we reduce the cutoff, the bare Lagrangian of the full theory,
which is specified by the boundary conditions (\ref{fbc}) and (\ref{fbc2}),
flows down to a submanifold parametrized by relevant couplings only.
It has both heavy and light operators. Now the flow may be projected
onto the submanifold of operators consisting
of the light field only. Then finding a local low-energy effective field theory
of light particles is equivalent to finding a renormalization group
flow with
a few (relevant or irrelevant) local operators which can
best approximate the projected flow of the full theory. To the zeroth order
in $1/M^2$, it is done with relevant terms only by choosing the same
renormalization conditions as those of the full theory. The deviation from the
full-theory flow is corrected by reading off values of the remaining irrelevant
coordinates at the $\LLL=0$ point. At first, components of dimension six
operators may be read, which tell us the effects of order $1/M^2$;
the flow of
the full theory is now approximated to the order $1/M^2$ at $\LLL<M$. If one
wants to increase the accuracy, it is necessary to read more and more
irrelevant coordinates (at $\LLL=0$) of the projected flow of the full
theory and modify the effective-theory flow with the corresponding
flow equation given in
Eq.\ (\ref{flowk}). Our equation (\ref{flowk}) automatically takes care of
possible divergences due to the unnaturally large irrelevant components
(see the discussion of Sec.\ 1) in such a way that the bare Lagrangian
may contain only a finite number of irrelevant terms.

\setcounter{equation}{0}
\section{CONCLUSIONS}
\ \indent
In the context of a simple scalar field theory, we have demonstrated that,
at low energy,
virtual heavy particle effects on light-particle Green functions are completely
factorized via effective local vertices to any desired order. For this, we
have used the powerful flow equation approach which can be regarded as
a differential version of Wilson's
renormalization group transformation. Remarkably, the method which has been
used previously to prove renormalizability is found to have an immediate
generalization to this problem. In
applying this method, we have seen that irrelevant terms in the Wilsonian
effective Lagrangian (which has a finite UV cutoff $\LLL=M$,
a characteristic
scale representing heavy-particle thresholds) are replaceable by
the corresponding higher
dimensional composite operators plus counterterms for their products in the
conventional approach (where UV cutoff is supposed to go to infinity).
The latter can be dealt with with the help of the normal product concept.
Once this fact is noticed, factorization is straightforward:
all $1/M^{2N}$-order effects can be isolated in terms of local vertices
involving operators of dimension $(4+2N)$. Thereby we arrive at
a local effective field theory which describes low-energy light-particle
physics accurately up to any desired order in $1/M^2$ with appropriate
calculation rules for irrelevant (nonrenormalizable) vertices.
Since the arguments here are essentially
dimensional, it should not be difficult to generalize them
to different field theories such as
gauge theories or theories with spontaneous symmetry breaking. As for these
models, we note that the
flow equation approach has already been used to discuss
renormalizability and related problems \cite{warr,keller,bonini}.

In this paper we limited our attention to vertex functions at momentum range
$\m=O(m)$.
As one increases $\m$ to sufficiently high energy scale,
the $(\m/m)$-dependence we neglected will become important.
Keeping such $\m$-dependence and establishing bounds more carefully, one
may study the high-momentum behavior of Green functions
(even in the exceptional
momentum region) within the effective field theory context.
In this regard, we note that the infrared behavior of massless
scalar theory has been analyzed by using
flow equations recently
\cite{keller,bonini}. Another
related point is the following. Effective theory
is supposed to be useful mainly at low energy.
As we increase energy, we need in fact
more and more irrelevant terms to maintain
the same level of accuracy. Conversely, with a few irrelevant terms added to
the Lagrangian, we may extend the energy range where it remains effective.
Then it
seems worth investigating quantitatively to what energy scale the given
effective theory may be used to reproduce the full-theory results
with reasonable accuracy.
\section*{NOTE ADDED}
\ \indent
Completing the main parts of this paper, we became aware of the very recent
paper by Girardello and Zaffaroni \cite{girardello} which has
a partial overlap with the present paper. They also discussed decoupling
and factorization to order $1/M^2$ using flow equations. However, we
believe that our treatment is physically more transparent and systematic.
For example, they did not discuss the normal product description
nor higher order generalization; this does not look trivial in their method
for, in their approach to these issues, nonnegligible parts of heavy-field
modes might remain unintegrated even at scales much less than the heavy
field mass $M$.
\section*{ACKNOWLEDGMENTS}
\ \indent
It is a great pleasure to thank Professor Choonkyu Lee for suggesting this
problem and carefully reading the manuscript. This work was supported in part
by the Korea Science and Engineering Foundation (through Center for
Theoretical Physics, SNU) and by the Basic Science Research Institute Program
(Project No.\ BSRI-94-2418), Ministry of Education, Korea.

\renewcommand{\theequation}{A.\arabic{equation}}
\setcounter{equation}{0}
\section*{APPENDIX}
\ \indent
Here we briefly discuss the renormalization of composite operators and their
products \cite{hughes,keller}, within the $\p^4$-theory defined by $\lilo$
in Eq.\ (\ref{cebare}).
We restrict our discussions to those needed in the main part of this
paper.
The operator $\lilon$'s of Sec.\ 2 are assumed to be of
local, even polynomials in $\p$ and its derivatives of dim.$\le 2D_N$,
\be \label{liloia}
\lilon=
  \sum_{n=1}^{D_N}\sum_{2n+|w|\le2D_N}
         b_{2n,\{w\}}^{N}\mono{2n}{w}\,,
\ee
where $b_{2n,\{w\}}^{N}$ is a formal power series
in $g_1$, $b_{2n,\{w\}}^{N}=\sum_{r=0}^\infty g_1^rb_{r;2n,\{w\}}^N$,
which is to be determined
later uniquely when we impose renormalization
conditions.
[$\mono{2n}{w}$ is
defined in Eq.\ (\ref{mono}).]
Throughout this Appendix, the dimension $D_N$ is restricted to satisfy the
condition $(2D_N-4)\ge\sum_{i=1}^kN_i(2D_i-4)$,
where $D_i\eq D_N|_{N_j=\d_{ij}}$, $N=(N_1,\ldots,N_k)$.
We follow the now familiar procedure. $\lvln{r}{2n}$'s
are defined by Eq.~(\ref{lilnexpansion}) and have
similar properties
except for the Euclidean invariance\footnote{Since the flow equation
respects $O(4)$ Euclidean symmetry, one may well restrict the consideration to
Euclidean-invariant operators and then $\lvln{r}{2n}$ will also be Euclidean
invariant.}.
They satisfy the flow equation (\ref{lvlnflow}).
Boundary conditions are given similarly:
\bea \label{lvlnbc1a}
\LLL=\lo:&
\pa_p^w\lvlon{r}{2n}=0,&\mbox{\hspace{10mm}for }2n+|w|>2D_N\,,\\
\LLL=0:&\hspace{3mm}\pa_p^w\lvon{r}{2n}(0)=b_{r;2n,\{w\}}^{R;N}\caln\,,
          & \mbox{\hspace{10mm}for }2n+|w|\le 2D_N\,,\label{lvlnbc2a}
\eea
where $\caln$ is a combinatorial factor defined by
$\left.\pa_p^{w'}\left[(ip_1)^{w_1}\cdots(ip_{2n-1})^{w_{2n-1}}\right]_{symm}
        \right|_{p=0}=\d_{\{w\}\{w'\}}\caln\,,
$
and
$b_{r;2n,\{w\}}^{R;N}$'s are finite and $\lo$-independent constants.
[Here it is to be noted that
$b_{2n,\{w\}}^N$ and $b_{2n,\{w\}}^{R;N}$ are in general not the same; this is
the usual operator mixing \cite{hughes,keller}.]
With these one can easily prove the perturbative renormalizability of Green
functions with the insertion of operator $O_N$ (which is defined in
Eq.\ (\ref{oiexpansion})); the proof is just a simpler version of that in
Theorem 4.

We now discuss Zimmermann's normal product \cite{zimmermann}, following
Ref.\ \cite{keller}\footnote{Here we are discussing only space-integrated
operators, or vertices.
Generalization to local operators, e.g. $[\mono{2n}{w}(x)]$, are
straightforward and is described in Ref.\ \cite{keller} to the
order $k=2$ and $|N|=2$.}.
We begin with a monomial for the case of $k= N=1$ with $D_1\eq D$.
Normal product $[\mono{2n'}{w'}]_{2D}$ of dimension $2D$ $(\ge 2n'+|w'|)$ is
defined by
requiring that the vertices with one insertion of
$O_1=[\mono{2n'}{w'}]_{2D}$, which are denoted as
$\lvlg{1}{r}{2n}([\mono{2n'}{w'}]_{2D})$,
obey the renormalization condition (\ref{lvlnbc2a}) with
\be \label{monobc}
b_{r;2n,\{w\}}^{R;1}=\d_{nn'}\d_{\{w\}\{w'\}}\d_{r0}\,.
\ee
Since the flow equation is linear, for general boundary condition
(\ref{lvlnbc2a}) we may
write $\lilog{1}$ as
\be \label{baseexpansion}
\lilog{1}=\sum_{2n+\{w\}\le2D}b_{2n,\{w\}}^{R;1}[\mono{2n}{w}]_{2D}
    \eq\left[\sum_{2n+\{w\}\le2D}b_{2n,\{w\}}^{R;1}\mono{2n}{w}\right]_{2D}\,.
\ee
When the dimension of the operator is equal to $2D$, we simply write
$[\cdots]_{2D}=[\cdots]$; for example,
$[\mono{2n}{w}]_{2D}=[\mono{2n}{w}]$ if $2D=2n+|w|$.
Therefore, if we choose $b_{2n,\{w\}}^{R;1}=0$ for $2n+|w|\neq 2D$, we may
omit the subscript $D$ in Eq.\ (\ref{baseexpansion}).
This is the case of
Sec.\ 4 (see Eq.\ (\ref{lilo1})).

The normal product of several operators (assuming $k>1$)
are defined inductively as follows.
Let us write $O_N=\lilon=[B_i]_{2D_i}$ for $N=(N_1,\ldots,N_k)$ with
$N_j=\d_{ij}$ (i.e., $|N|=1$). Suppose that
$[[B_{i_1}]_{2D_{i_1}}\cdots[B_{i_{k-1}}]_{2D_{i_{k-1}}}]_{2D}$'s have been
defined, where $(2D-4)\ge\sum_{l=1}^{k-1}(2D_{i_l}-4),$ $1\le i_l\le k$.
Then we define the normal product
$[[B_1]_{2D_1}\cdots[B_k]_{2D_k}]_{2D_N}$ as $O_{(1,\ldots,1)}$ which flows
down through Eq.\ (\ref{flowk}) with the simplest boundary condition
\be \label{multiplebc}
\pa_p^w\lvog{(1,\ldots,1)}{r}{2n}(0)
         =b_{r;2n,\{w\}}^{R;(1,\ldots,1)}\caln=0\,,\quad 2n+|w|\le 2D_N\,.
\ee
If $(2D_N-4)=\sum_i(2D_i-4)$, we write $[\cdots]_{2D_N}\eq[\cdots]$
as above. Let us illustrate this definition for $k=2$ and $k=3$ cases.
If $k=2$, $[[B_1]_{2D_1}[B_2]_{2D_2}]_{2D_{(1,1)}}$ is defined as
\be \label{b1b2}
[[B_1]_{2D_1}[B_2]_{2D_2}]_{2D_{(1,1)}}\eq O_{(1,1)}
      =[B_1]_{2D_1}[B_2]_{2D_2}+\lilog{(1,1)}\,,
\ee
where the flow of $\lilog{(1,1)}$ is governed by Eq.\ (\ref{flowk}) together
with the boundary
condition (\ref{multiplebc}). Therefore, $\lilog{(1,1)}$ is interpreted as
the counterterm
needed to cancel divergences appearing in Green functions with the
insertion of the product operator $[B_1]_{2D_1}[B_2]_{2D_2}$.
Similarly, if $k=3$,
from the
expression for $O_{(1,1,1)}$ given in Eq.\ (\ref{oi}) we have
\bea
[[B_1]_{2D_1}[B_2]_{2D_2}[B_3]_{2D_3}]_{2D_{(1,1,1)}}
     &\eq&O_{(1,1,1)}\\
 &=&[B_1]_{2D_1}[B_2]_{2D_2}[B_3]_{2D_3}+[B_1]_{2D_1}\lilog{(0,1,1)}\no\\
  &&+[B_2]_{2D_2}\lilog{(1,0,1)}+[B_3]_{2D_3}\lilog{(1,1,0)}
  +\lilog{(1,1,1)}\,,\no
\eea
where $\lilog{(1,1,0)}$ etc.\ are given by Eq.\ (\ref{b1b2}),
and additional divergences due to multiple insertions are cancelled
by $\lilog{(1,1,1)}$ which obeys
the flow equation (\ref{flowk}) with $N=(1,1,1)$, with the boundary condition
(\ref{multiplebc}). In this way, $[[B_1]_{2D_1}\cdots[B_k]_{2D_k}]_{2D_N}$ is
defined for general $k$, i.e., $\lilog{(1,\ldots,1)}$ serves as the
``counterterm'' for new divergences appearing in Green functions due to the new
products of lower order operators and $[[B_1]_{2D_1}\cdots[B_k]_{2D_k}]_{2D_N}$
is interpreted as the resulting ``subtracted'' finite part.
Since this definition satisfies multilinearity,
which can be proved by use of induction on $k$ while taking into
account Eqs.\ (\ref{oi}) and (\ref{flowk}), and boundary
conditions, we can freely put
the coefficients multiplying operators inside or outside the brackets.

For the general boundary conditions (\ref{lvlnbc2a}) where $b_{2n,\{w\}}^{R;N}$
need not be zero, we recall the property of the flow equation (\ref{flowk})
mentioned at the end of Sec.\ 2. The general solution of the flow equation is
represented as a sum of a particular solution and a solution of the
homogeneous equation which is identical to the flow equation with $|N|=1$,
i.e.,
\be \label{inhomogeneous}
\liln=\lilg{1}([B_N]_{2D_N})+\liln_{CT}\,.
\ee
Here we may absorb all the nonzero boundary values into
$\lilog{1}([B_N]_{2D_N})$ choosing
\be \label{bik1}
[B_N]_{2D_N}=\sum_{2n+|w|\le 2D_N}b_{2n,\{w\}}^{R;N}[\mono{2n}{w}]_{2D_N}\,,
\ee
and then $\liln_{CT}$ satisfies the simplest boundary condition
(\ref{multiplebc}).
(The subscript $CT$ is attached to indicate its counterterm nature as
discussed above.) In other words, $b_{2n,\{w\}}^{R;N}\neq0$
corresponds
to adding new composite operator $[B_N]_{2D_N}$ to $\lilon$
besides the counterterms needed just to
cancel divergences associated with the
multiple insertion of operators already included.
For example,if $k=2$, we may extend
Eq.\ (\ref{b1b2}) to the general case,
\bea \label{b1b2general}
\lilog{(1,1)}
 &=&\lilog{(1,1)}_{CT}+[B_{(1,1)}]_{2D_{(1,1)}}\\
 &=&[[B_1]_{2D_1}[B_2]_{2D_2}]_{2D_{(1,1)}}-[B_1]_{2D_1}[B_2]_{2D_2}
   +\sum_{2n+|w|\le 2D_{(1,1)}}b_{2n,\{w\}}^{R;(1,1)}
                      [\mono{2n}{w}]_{2D_{(1,1)}}\,.\no
\eea
This is the case we encountered in Sec. 4.

\renewcommand{\np}[1]{{\it Nucl.\ Phys.\ B {\bf #1}}}
\renewcommand{\plt}[1]{{\it Phys.\ Lett.\ B {\bf #1}}}
\newcommand{\prd}[1]{{\it Phys.\ Rev.\ D {\bf #1}}}
\newcommand{\prb}[1]{{\it Phys.\ Rev.\ B {\bf #1}}}
\renewcommand{\prlt}[1]{{\it Phys.\ Rev.\ Lett.\ {\bf #1}}}
\renewcommand{\rmps}[1]{{\it Rev.\ Mod.\ Phys.\ {\bf #1}}}
\renewcommand{\prp}[1]{{\it Phys.\ Rep.\ {\bf #1}}}
\renewcommand{\anp}[1]{{\it Ann.\ Phys. (N.\ Y.)\ {\bf #1}}}
\renewcommand{\cmp}[1]{{\it Comm.\ Math.\ Phys.\ {\bf #1}}}
\renewcommand{\ijmp}[1]{{\it Int.\ J.\ Mod.\ Phys.\ A {\bf #1}}}
\def\mybibliography#1{\begin{center}\subsection*{{\bf REFERENCES}%
}\end{center}\list
  {\arabic{enumi}.}{\settowidth\labelwidth{[#1]}
\leftmargin\labelwidth    \advance\leftmargin\labelsep
    \usecounter{enumi}}
    \def\newblock{\hskip .11em plus .33em minus .07em}
    \sloppy\clubpenalty4000\widowpenalty4000
    \sfcode`\.=1000\relax}

\let\endmybibliography=\endlist

\begin{mybibliography}{99}
\bibitem{wilson}K. G. Wilson, \prb{4} (1971), 3174, 3184.
\bibitem{polchinski}J. Polchinski, \np{231} (1984), 269.
\bibitem{warr}B. J. Warr, \anp{183} (1988), 1, 89.
\bibitem{hughes}J. Hughes and J. Liu, \np{161} (1988), 183.
\bibitem{keller}G. Keller, C. Kopper and M. Salmhofer, Helv. Phys. Acta
 {\bf 65} (1992), 32; G. Keller and C. Kopper, \plt{273} (1991), 323;
 \cmp{148} (1992), 445; {\bf 153} (1993), 245; \cmp{161} (1994), 515.
\bibitem{bonini}M.\,Bonini, M.\,D'Attanosio and G.\,Marchesini,
    \np{409}\,(1993),441; 418\,(1994),81.
\bibitem{wetterich}C. Wetterich, \plt{301} (1993), 90 and references therein.
\bibitem{ellwanger} U. Ellwanger, {\it Z.\ Phys.\ C} {\bf 62} (1994), 503.
\bibitem{morris}T. R. Morris, \ijmp{9} (1994,2411.
\bibitem{ball}R. D. Ball and R. S. Thorne, CERN-TH.7067/93, OUTP-93-23P.
\bibitem{applequist}T. Appelquist and J. Carazzone, \prd{11} (1975), 2856.
\bibitem{weinberg}S. Weinberg, \plt{91} (1980), 51.
\bibitem{clee}C. Lee, \np{161} (1979), 171.
\bibitem{kazama}Y. Kazama and Y. P. Yao, \prd{21} (1980), 1116, 1138.
\bibitem{zimmermann}W. Zimmermann, {\it in} ``Lectures on Elmentary
Particles and Fields'', (S. Deser, M. Grisaru and H. Pendleton, Eds)
MIT Press, Cambridge, MA, 1970.
\bibitem{symanzik}K. Symanzik, \np{226} (1983), 187, 205.
\bibitem{keller4}C. Wieczerkowski, \cmp{120} (1988), 148;
G. Keller, Helv. Phys. Acta {\bf 66} (1993), 453.
\bibitem{collins}See for example J. C. Collins, ``Renormalization'',
Cambridge University press, 1984.
\bibitem{ovrut} See for example B.\ Ovrut and H.\ Schnitzer, \prd{22} (1980),
3369; {\it D\/} {\bf 22} (1980), 2518; H.\ Georgi, {\it Nucl.\ Phys.\
\/}(Proc.\ Suppl.) 29B,C (1992), 1.
\bibitem{girardello}L. Girardello and A. Zaffaroni, \np{424} (1994), 219.
\end{mybibliography}

\newpage
\section*{Figure caption}
\ \indent
Fig.\ 1. Graphical representation of the flow equation (\ref{lflflow}).
Thin (thick) lines represents light (heavy) field, and numerical factors in
Eq.\ (\ref{lflflow}) are ignored here.
\end{document}